\documentclass[11pt]{article}

\usepackage{float}
\usepackage{color}
\usepackage{tikz}
\usepackage{setspace}
\usepackage[toc,page]{appendix}
\usepackage{amssymb}
\usepackage{showlabels}
\usepackage[normalem]{ulem}
\usepackage{multirow}
\usepackage{tikz}
\usepackage[toc]{appendix}
\usepackage{chngcntr}
\usetikzlibrary{shapes,arrows}
\usepackage{subfigure}
\usepackage{authblk}
\usepackage{comment}
\usepackage{placeins}
\usepackage{cancel}
\usepackage{ulem}

\usepackage[round]{natbib}
\usepackage{latexsym}
\usepackage{amsmath}
\usepackage{graphicx}
\usepackage{caption}

\usepackage{bm}
\usepackage[pdftex,hypertexnames=false,linktocpage=true]{hyperref}
\hypersetup{colorlinks=true,linkcolor=blue,anchorcolor=blue,citecolor=blue,filecolor=blue,urlcolor=blue,bookmarksnumbered=true,pdfview=FitB}

\newcommand{\biblist}{\begin{list}{}
		{\listparindent 0.0cm \leftmargin 0.50cm \itemindent -0.50 cm
			\labelwidth 0 cm \labelsep 0.50 cm
			\usecounter{list}}\clubpanelty4000\widowpanelty4000}
	\newcommand{\ebiblist}{\end{list}}

\newcommand{\bx}{\boldsymbol{x}}

\newcommand{\bz}{\boldsymbol{z}}

\newcommand{\bg}{\mathbf{g}}

\newcommand{\bX}{\boldsymbol{X}}

\newcommand{\btheta}{\boldsymbol{\theta}}
\newcommand{\bxi}{\boldsymbol{\xi}}

\newcommand{\bgamma}{\boldsymbol{\gamma}}

\newcommand{\bpsi}{\boldsymbol{\psi}}

\begin{document}
	
\def\spacingset#1{\renewcommand{\baselinestretch}%
	{#1}\small\normalsize} \spacingset{1}


	\title{\bf Maximum entropy classification\\ for record linkage}
    \author[1]{Danhyang Lee}
    \author[2,3,4]{Li-Chun Zhang}
    \author[5]{Jae Kwang Kim}
    \affil[1]{Department of Information Systems, Statistics and Management Science, University of Alabama, Tuscaloosa, AL, U.S.A.}
    \affil[2]{Department of Social Statistics and Demography, University of Southampton, Southampton, U.K.}
    \affil[3]{Statistics Norway, Oslo, Norway}
    \affil[4]{Department of Mathematics, University of Oslo, Oslo, Norway}
    \affil[5]{Department of Statistics, Iowa State University, Ames, IA, U.S.A.}
	\maketitle


\bigskip

\begin{abstract}
	By record linkage one joins records residing in separate files which are believed to be related to the same entity. In this paper we approach record linkage as a classification problem, and adapt the maximum entropy classification method in {machine learning} to record linkage, both in the supervised and unsupervised settings of machine learning. The set of links will be chosen according to the associated uncertainty. {On the one hand, our framework overcomes some persistent theoretical flaws of the classical approach pioneered by \cite{fellegi1969theory}; on the other hand, the proposed algorithm is fully automatic, unlike the classical approach that generally requires clerical review to resolve the undecided cases. }
\end{abstract}

\noindent%
\vfill

\newpage
\spacingset{1.45} 

\section{Introduction}
\label{sec:intro}

Combining information from multiple sources of data is a frequently encountered problem in many disciplines. To combine information from different sources, one assumes that it is possible to identify the records associated with the same entity, which is not always the case in practice. {{The entity may be individual, company, crime, etc.}} If the data do not contain unique identification number, identifying records from the same entity becomes a challenging problem. \emph{Record linkage} is the term describing the process of joining records that are believed to be related to the same entity. While record linkage may entail the linking of records within a single computer file to identify duplicate records, referred to as \emph{deduplication}, we focus on linking of records across separate files.

{Record linkage (RL) has been employed for several decades in survey sampling producing official statistics. In particular, linking administrative files with survey sample data can greatly improve the quality and resolution of the official statistics. As applications, \cite{jaro1989advances} and  \cite{winkler1991application} merged post-enumeration survey and census data for census coverage evaluation. \cite{zhang2012data} linked population census data files over time, and \cite{Owen2015large} linked administrative registers to create a single statistical population dataset. The classical approach pioneered by \cite{fellegi1969theory}, which is the most popular method of RL in practice, has been successfully employed for these applications.} 

The probabilistic decision rule of \cite{fellegi1969theory} is based on the likelihood ratio test idea, 
by which we can determine how likely a particular record pair is a true match. In applying the likelihood ratio test idea, one needs to estimate the model parameters of the underlying model and determine the thresholds of the decision rule. {\cite{winkler1988} and \cite{jaro1989advances} treat the matching status as an unobserved variable and propose an EM algorithm for parameter estimation, which we shall refer to as the WJ-procedure}. See \cite{herzog2007data}, \cite{christen2011survey} and \cite{binette2020almost} for overviews. {However, as explained in Section \ref{sec:problems}, to motivate the WJ-procedure as an EM algorithm requires the crucial assumption that measures of agreement between the record pairs, called \textsl{comparison vectors}, are independent from one record pair to another, which is impossible to hold in reality. \cite{newcombe1959automatic} address dependence between comparison vectors through data application. Also, see e.g. \cite{tancredi2011hierarchical}, \cite{sadinle2017bayesian}, and \cite{binette2020almost} for discussions of this issue.} {Bayesian approaches to RL are also available in the literature \citep{steorts2015, sadinle2017bayesian, stringham2021}. 
Bayesian approaches to RL problems  allow us to quantify uncertainty on
the matching decisions. However, the stochastic search using MCMC algorithm in the Bayesian approach involves extra computational burden.  }

To develop an alternative approach, we first note that the RL  problem is essentially a classification problem, where each record pair is classified into either ``match'' or ``non-match'' class.  Various classification techniques based on machine learning approaches have been employed for record linkage \citep{hand2018note, christen2011survey,christen2008automatic, sarawagi2002interactive}. In this paper, we adapt the maximum entropy method for {classification} to record linkage. Specifically, we can view the likelihood ratio of the method proposed by \cite{fellegi1969theory} as a special case of the density ratio and apply the maximum entropy method for density ratio estimation. For example, \cite{nigam1999using} use the maximum entropy for text classification and \cite{nguyen2010estimating} develop a more unified theory of maximum entropy method for density ratio estimation. {There is, however, a key difference of record linkage to the standard setting of classification problems, in that the different record pairs are not distinct `units' because the same record is part of many record pairs.}

We present our maximum entropy record linkage algorithm for both supervised and unsupervised settings, while our main contributions concern the unsupervised case. Supervised approaches need training data, i.e., record pairs with known true match and true non-match status. Such training data are often not available in real world situations, or have to be prepared manually, which is very expensive and time-consuming \citep{christen2007two}. {Thus, the unsupervised case is by far the most common in practice.} In the unsupervised case, however, one cannot estimate the density ratio directly based on the observed true matches and non-matches, and it is troublesome to jointly model for the unobserved match status and the observed comparison scores over all the record pairs. {We develop a new iterative algorithm to jointly estimate the density ratio as well as the maximum entropy classification set in the unsupervised setting and prove its convergence. The associated measures of the linkage uncertainty are also developed.}


Furthermore, we show that the WJ-procedure can be incorporated as a special case of our approach to estimation,
but without the need of the independence assumption between the record pairs. This reveals that the WJ-procedure can be motivated without the 
independence assumption, and explains why it gives reasonable results in many situations. The choice of the set of links is guided by the uncertainty measures developed in this paper. This is an important practical improvement over the classical approach, which does not directly provide any uncertainty measure for the final set of links. Our procedure is fully automatic, without the need for resource-demanding clerical review that is required under the classical approach. 

The paper is organised as follows. In Section \ref{sec:problems}, the basic setup and the classical approach are introduced. In Section \ref{sec:MECsup}, the proposed method is developed under the setting of supervised record linkage. In Section \ref{sec:MECunsup}, we extend the proposed method to the more challenging case of the unsupervised record linkage. Discussions of some related estimation approaches and technical details are presented in Section \ref{sec:discussion} and Appendix. Results from an extensive simulation study are presented in Section \ref{sec:simul}. Some concluding remarks {and comments on further works} are given in Section \ref{sec:final}. 

\section{Problems with the classical approach} 
\label{sec:problems}

Suppose that we have two data files $A$ and $B$ that are believed to have many common entities but no duplicates within each file. Any record in $A$ and another one in $B$ may or may not refer to the same entity. Our goal is to find the true matches among all possible pairs of the two data files. Let the bipartite \emph{comparison space} $\Omega = A\times B = M\cup U$ consist of \emph{matches} $M$ and \emph{non-matches} $U$ between the records in files $A$ and $B$. For any pair of records $(a,b) \in \Omega$, let $\bgamma_{ab}$ be the \emph{comparison vector} between a set of \emph{key} variables associated with $a\in A$ and $b\in B$, respectively, such as name, sex, {date of birth}. The key variables and the comparison vector $\bgamma_{ab}$ are fully observed over $\Omega$. In cases where the key variables may be affected by errors, a match $(a,b)$ may not have complete agreement in terms of $\bgamma_{ab}$, and a non-match $(a,b)$ can nevertheless agree on some (even all) of the key variables. 

In the classical approach of \cite{fellegi1969theory}, one recognizes the probabilistic nature of $\bgamma_{ab}$ due to the perturbations that cause key-variable errors. The related methods are referred to as \emph{probabilistic record linkage}. 
{To explain the probabilistic record linkage method of \cite{fellegi1969theory}, let $m(\bgamma_{ab})$ $= f(\bgamma_{ab} \mid (a,b) \in M )$ be the probability mass function of the discrete values $\bgamma_{ab}$ can take given $(a,b)\in M$. Similarly, we can define $u(\bgamma_{ab})$ $=  f(\bgamma_{ab} \mid (a,b) \in U)$.} The ratio 
\[
r_{ab} = {\frac{m(\bgamma_{ab})}{u(\bgamma_{ab})}} 
\]
is then the basis of the likelihood ratio test (LRT) for $H_0: (a,b)\in M$ vs. $H_1: (a,b)\in U$. Let $M^* = \{ (a,b) : r_{ab} > c_M \}$ be the pairs classified as matches and $U^* = \{ (a,b) : r_{ab} < c_U \}$ the non-matches, the remaining pairs are classified by clerical review, where $(c_M, c_U)$ are {the thresholds} related to the probabilities of false links (of pairs in $U$) and false non-links (of pairs in $M$), respectively, defined as
\begin{equation} \label{FS}
\mu = \sum_{\bgamma} u(\bgamma) \delta(M^* ; \bgamma) \qquad\text{and}\qquad 
\lambda = \sum_{\bgamma} m(\bgamma) \delta(U^* ; \bgamma),
\end{equation}
where $\delta(M^* ; \bgamma) =1$ if $\bgamma_{ab} = \bgamma$ means $(a,b)\in M^*$ and 0 otherwise, similarly for $\delta(U^* ; \bgamma)$.

In practice the probabilities $m(\bgamma)$ and $u(\bgamma)$ are unknown. Neither is the \emph{prevalence} of true matches, given by $\pi = |M|/|\Omega| := n_M/n$. Let $\boldsymbol{\eta}$ {be the set containing} $\pi$ and the unknown parameters of $m(\bgamma)$ and $u(\bgamma)$. 
Let $g_{ab} = 1$ if $(a,b)\in M$ and 0 if $(a,b)\in U$. Given the complete data $\{ (g_{ab}, \bgamma_{ab}) : (a,b)\in \Omega \}$, {\cite{winkler1988} and }\cite{jaro1989advances} assume the log-likelihood to be
\begin{equation} \label{jaro}
h(\boldsymbol{\eta}) = \sum_{(a,b)\in \Omega} g_{ab} \log (\pi {m(\bgamma_{ab})}) +  \sum_{(a,b)\in \Omega} (1-g_{ab}) \log \big( (1-\pi) {u(\bgamma_{ab})} \big). 
\end{equation}
An EM-algorithm follows by treating $g_{\Omega} = \{ g_{ab} : (a,b)\in \Omega\}$ as the missing data. 

There are two fundamental  problems with this classical approach. 
\begin{description} 
\item{[Problem-I]} Record linkage is not a direct application of the LRT, because one needs to evaluate \emph{all} the pairs in $\Omega$ instead of any \emph{given} pair. The classification of $\Omega$ into $M^*$ and $U^*$ is incoherent generally, since a given record can belong to multiple pairs in $M^*$. Post-classification deduplication of $M^*$ would be necessary then, {which is \emph{not} part of the theoretical formulation above. In particular, there lacks an associated method for estimating the uncertainty surrounding the final linked set, such as the amount of false links in it or the remaining matches outside of it.}       

\item{[Problem-II]}  In reality the comparison vectors of any two pairs are not independent, as long as they share a record. For example, given $(a,b)\in M$ and $\bgamma_{ab}$ not subjected to errors, then $g_{ab'}$ must be 0, for $b'\neq b$ and $b'\in B$, as long as there are no duplicated records in either $A$ or $B$, and $\bgamma_{ab'}$ depends only on the key-variable errors of $b'$. Whereas, marginally, $g_{ab'} =1$ with probability $\pi$ and $\bgamma_{ab'}$ depends also on the key-variable errors of $a$. It follows that $h(\boldsymbol{\eta})$ in \eqref{jaro} does not correspond to the true joint-data distribution of $\bgamma_{\Omega} = \{ \bgamma_{ab} : (a,b)\in \Omega\}$, even when the marginal $m$ and $u$-probabilities are correctly specified. Similarly, although one may define \emph{marginally} $\pi = \mbox{Pr}[ (a,b)\in M | (a,b)\in \Omega]$ for a \emph{randomly} selected record pair from $\Omega$, it does not follow that $\log f(g_{\Omega}) = n_M \log\pi + (n - n_M) \log (1-\pi)$ \emph{jointly} as in \eqref{jaro}. For both reasons, $h(\boldsymbol{\eta})$ given by \eqref{jaro} cannot be the complete-data log-likelihood. 
\end{description} 

In the next two sections, we develop maximum entropy classification to record linkage to {avoid the problems above}, after which more discussions of the classical approach will be given.

\section{Maximum entropy classification: Supervised} 
\label{sec:MECsup}

As noted in Section 1, {the} record linkage problem is a  classification problem. Maximum entropy classification has been used in image restoration or text analysis (\citealp{gull1984, Berger1996}). 
 \emph{Maximum entropy classification (MEC)} has been proposed for supervised learning (SL) to standard classification problems, where the units are known but the true classes of the units are unknown apart from a sample of \emph{labelled units}. Let $Y \in \{ 1, 0\}$ be the true class and $\bX$ the random vector of features. Let the density ratio be
\[
r(\boldsymbol{x}; \boldsymbol{\eta}) = \frac{f(\bx | Y=1; \boldsymbol{\eta}) }{ f(\bx | Y = 0; \boldsymbol{\eta}) } { :=  \frac{f_1(\bx ;  \boldsymbol{\eta}) }{ f_0(\bx ; \boldsymbol{\eta}) } ,}
\]
where $f_1$ and $f_0$ are the conditional density functions given $Y =1$ or $0$, respectively, and $\boldsymbol{\eta}$ contains the unknown parameters. For MEC based on $r(\bx)$, one finds $\hat{\boldsymbol{\eta}}$ that maximises the Kullback-Leibler (KL) divergence from $f_0$ to $f_1$ subjected a constraint, i.e.
\[
D = \int_{\mathcal{S}_1} f_1(\bx; \boldsymbol{\eta}) \log r(\bx; \boldsymbol{\eta}) d\bx \quad\text{subjected to}\quad
\int_{\mathcal{S}_1} f_0(\bx; \hat{\boldsymbol{\eta}}) r(\bx; \hat{\boldsymbol{\eta}}) d\bx = 1, 
\]
where $\mathcal{S}_1$ is the support of $\bX$ given $Y=1$, and the normalisation constraint arises
since $r(\bx; \hat{\boldsymbol{\eta}}) f_0(\bx;\hat{\boldsymbol{\eta}})$ is an estimate of $f_1(\bx)$. Provided common support $\mathcal{S}_1 = \mathcal{S}_0$, where {$\mathcal{S}_0$ is the support of $\bX$ given $Y=0$}, one can use the empirical distribution function (EDF) of $X$ over $\{ \bx_i : y_i = 1\}$ in place of $f_1$ for $D$, and that over $\{ \bx_i : y_i = 0\}$ in place of $f_0$ for the constraint. Having obtained $\hat{r}_{\bx} = r(\bx; \hat{\boldsymbol{\eta}})$, one can classify any unit given the associated feature vector $\bx$ based on $\mbox{Pr}(Y=1 | \bx; \hat{p}, \hat{r}_{\bx})$, where $\hat{p}$ is an estimate of the prevalence $p = \mbox{Pr}(Y=1)$.

We describe how the idea of MEC for supervised learning can be adapted to record linkage problem in the following subsections.

\subsection{Probability ratio for record linkage} 

For supervised learning based MEC to record linkage, suppose $M$ is observed for the given $\Omega$, and the trained classifier is to be applied to the record pairs outside of $\Omega$. To fix the idea, suppose $B$ is a non-probability sample that overlaps with the population $\mathcal{P}$, and $A$ is a probability sample from $\mathcal{P}$ with known inclusion probabilities. 
While $\bgamma_M =\{ \bgamma_{ab} : (a,b)\in M\}$ may be considered as an IID sample, since each $(a,b)$ in $M$ refers to a distinct entity, this is not the case with $\{ \bgamma_{ab} : (a,b)\not\in M \}$, whose \emph{joint} distribution is troublesome to model.

\paragraph{\textnormal{\em Probability ratio (I)}} Let $r_q(\bgamma)$ be the \emph{probability ratio} given by
\[
 r_q(\bgamma)= \frac{m(\bgamma) }{ q(\bgamma) }  , 
\]
where $m(\bgamma)$ is the probability mass function of $\bgamma_{ab} = \bgamma$ given $g_{ab}=1$, and $q(\bgamma)$ is that over $\bgamma_{\Omega} = \{ \bgamma_{ab} : (a,b) \in \Omega\}$. The KL divergence measure from $q(\bgamma)$ to $m(\bgamma)$ and the normalisation constraint are 
\[
D_f = \sum_{\bgamma\in \mathcal{S}(M)} m(\bgamma) \log {r_q(\bgamma)} \quad\text{and}\quad
\sum_{\bgamma \in \mathcal{S}(M)} {\hat{q}(\bgamma) \hat{r}_q(\bgamma)} =1 ~,
\]
where $\mathcal{S}(M)$ is the support of $\bgamma_{ab}$ given $g_{ab}=1$. 
This set-up allows $\mathcal{S}(M)$ to be a subset of $\mathcal{S}$, where $\mathcal{S}$ is the support of all possible $\bgamma_{ab}$. 
It follows that, based on the IID sample $\bgamma_{M}$ of size $n_M = |M|$, the objective function to be \emph{minimized} for {$r_q$} can be given by
\begin{equation} \label{Qf-SL}
Q_f = \sum_{(a,b)\in M} \frac{f(\bgamma_{ab})}{n_M(\bgamma_{ab})} {r_q(\bgamma_{ab})} 
- \frac{1}{n_M} \sum_{(a,b)\in M} \log {r_q(\bgamma_{ab})}, 
\end{equation}
where $n_M(\bgamma_{ab}) = \sum_{(i,j)\in M} \mathbb{I}(\bgamma_{ij} = \bgamma_{ab})$ based on the observed support $\mathcal{S}(M)$.

\paragraph{\textnormal{\em Probability ratio (II)}}  \emph{Provided} $\mathcal{S}(M) \subseteq \mathcal{S}(U)$, where $\mathcal{S}(U)$ is the support of $\bgamma_{ab}$ over $U$, one can let the probability ratio be given by
\[
r(\bgamma) = \frac{m(\bgamma)}{ u(\bgamma) } 
\]
where $u(\bgamma)$ is the probability of $\bgamma_{ab} = \bgamma$ given $g_{ab} = 0$. We have 
\[
{r_q(\bgamma)} = \frac{ m( \bgamma)}{ {q( \bgamma)} } = \frac{ m(\bgamma) }{ \pi m( \bgamma) + (1-\pi) u( \bgamma) } 
= \frac{r(\bgamma)}{ \pi \big( r( \bgamma) -1\big) + 1} 
\]
where ${q(\bgamma)} = \pi m( \bgamma) + (1-\pi) u( \bgamma)$, so that ${r_q(\bgamma)}$ and $r(\bgamma)$ are one-to-one. Meanwhile, the KL divergence measure from $u(\bgamma)$ to $m(\bgamma)$ is given by
\[
D = \sum_{ \bgamma \in \mathcal{S}(M)} m( \bgamma) \log r( \bgamma) 
\]
and the objective function to be \emph{minimized} for $r$ can now be given by
\begin{align}
& Q = \sum_{(a,b)\in M} \frac{u(\bgamma_{ab})}{n_M(\bgamma_{ab})} r(\bgamma_{ab}) 
- \frac{1}{n_M} \sum_{(a,b)\in M} \log r(\bgamma_{ab}). \label{Q-SL} 
\end{align}

\paragraph{\textnormal{\em Models of $\bgamma$}:} Under the multinomial model, one can simply use the EDF of $\bgamma$ over $\bgamma_{\Omega}$ as $f(\bgamma)$, for each distinct level of $\bgamma$, as long as $|\Omega|$ is large compared to $|\mathcal{S}|$. Similarly for $m(\bgamma)$ over $\bgamma_M$ and $u(\bgamma)$ over $U$. For linkage outside of $\Omega$, the estimated $m(\bgamma)$ from $M(\Omega)$ applies, if the selection of $A$ from $\mathcal{P}$ is non-informative.

For $\bgamma$ made up of $K$ binary agreement indicators, $\gamma_k = 0,1$ for $k=1, \ldots, K$, there are up to $2^K$ distinct levels of $\bgamma$, which can sometimes be relatively large compared to $|M|$. A more parsimonious model of $m(\bgamma; \btheta)$ that is commonly used is given by
\begin{equation} \label{m-model}
m(\bgamma; \btheta) = \prod_{k=1}^K \theta_k^{\gamma_k} (1- \theta_k)^{1- \gamma_k} 
\end{equation}
where $\theta_k = \mbox{Pr}(\gamma_{ab,k} = 1 | g_{ab} = 1)$, and $\gamma_{ab,k}$ is the $k$-th component of $\bgamma_{ab}$. {It is possible to model $\theta_k$ based on the distributions of the key variables that give rise to $\bgamma$, which makes use of the differential frequencies of their values, such as the fact that some names are more common than others.} {Similarly, $u(\bgamma; \boldsymbol{\xi})$ can be modeled as in (\ref{m-model}) with parameters $\xi_k$ instead of $\theta_k$, where $\xi_k = Pr(\gamma_{ab,k}=1 \mid g_{ab}=0)$.}

{Note that (\ref{m-model}) implies conditional independence among agreement indicators. \citet{winkler1993} and \citet{winkler1994} demonstrated that even when the conditional independence assumption does not hold, results based on conditional independence assumption are quite robust.} More complicated models that allow for correlated $\gamma_k$ can also be considered. {See \cite{armstrong1993model} and \cite{larsen2001iterative} for discussion of those models.} {See \cite{xu2019incorporating} for a recent study which compares models with or without correlated $\gamma_k$.}

\subsection{MEC sets for record linkage} \label{MEC-SL}

{Provided there are no duplicated records in either $A$ or $B$},  a \emph{classification set} for record linkage, denoted by $\hat{M}$, consists of record pairs from $\Omega$, where any record in $A$ or $B$ appears at most in one record pair in $\hat{M}$.
Let the \emph{entropy} of a classification set $\hat{M}$ be given by
\begin{eqnarray}
D_{\hat{M}} = \frac{1}{|\hat{M}|} \sum_{(a,b) \in \hat{M}} \log r(\bgamma_{ab}) \label{ent}
\end{eqnarray}
A MEC set of given size $n^* = |\hat{M}|$ is the first classification set that is of size $n^*$, obtained by deduplication in the descending order of $r(\bgamma_{ab})$ over $\Omega$. It is possible to have $(a,b')\not \in \hat{M}$ and $r(\bgamma_{ab'}) > r(\bgamma_{a',b'})$ for $(a',b')\in \hat{M}$, if there exists $(a,b) \in \hat{M}$ with $r(\bgamma_{ab}) > r(\bgamma_{ab'})$. 

A MEC set of size $n^*$ is not necessarily the largest possible classification set with the maximum entropy, to be referred to as a \emph{maximal} MEC set, which is the largest classification set such that $r(\bgamma_{ab}) = \max_{\bgamma} r(\bgamma)$ for every $(a,b)$ in it. In practice, a maximal MEC set is given by the first pass of \emph{deterministic linkage}, which only consists of the record pairs with perfect \emph{and} unique agreement of all the key variables. 

Probabilistic linkage methods for MEC set are useful if one would like to allow for additional links, even though their key variables do not agree perfectly with each other.  For the uncertainty associated with a given MEC set $\hat{M}$, we consider two types of errors. First, we define the \emph{false link rate (FLR)} among the links in $\hat{M}$ to be
\begin{equation} \label{psi} 
\psi = \frac{1}{|\hat{M}|} \sum_{(a,b)\in \hat{M}} (1- g_{ab})
\end{equation} 
which is different to $\mu$ by \eqref{FS} where the denominator is $|U|$. Second, the \emph{missing match rate (MMR)} of $\hat{M}$, which is related to the false non-link probability $\lambda$ in \eqref{FS}, is given by
\begin{equation} \label{tau}
\tau = 1 - \frac{1}{n_M} \sum_{(a,b)\in \hat{M}} g_{ab} . 
\end{equation}
While $\mu$ and $\lambda$ in \eqref{FS} are theoretical probabilities, the FLR and MMR are actual errors.

It is instructive to consider the situation, where one is asked to form MEC sets in $\Omega$ given all the necessary estimates related to the probability ratio $r(\bgamma)$, which can be obtained under the SL setting, without being given $n_M$, $g_{\Omega}$ or $M$ directly. 

First, the perfect MEC set should have the size $n_M$. Let $n(\bgamma) = \sum_{(a,b)\in \Omega} \mathbb{I}(\bgamma_{ab} = \bgamma)$. One can obtain $n_M$ as the solution to the following fixed-point equation: 
\begin{equation} \label{nM}
n_M = \sum_{(a,b)\in \Omega} \hat{g}(\bgamma_{ab}) = \sum_{\gamma\in \mathcal{S}} n(\bgamma) \hat{g}(\bgamma) 
\end{equation}
where 
\begin{equation} \label{ghat}
\hat{g}(\bgamma) := \mbox{Pr}(g_{ab} =1 | \bgamma_{ab} =\bgamma)
= \frac{\pi r(\bgamma)}{ \pi \big( r( \bgamma) -1\big) + 1} = \frac{n_M r(\bgamma)}{ n_M \big( r(\bgamma) -1 \big) + n} 
\end{equation}
and the probability is defined with respect to completely random sampling of a single record pair from $\Omega$. To see that $\hat{g}(\bgamma)$ by \eqref{ghat} satisfies \eqref{nM}, notice $\hat{g}(\bgamma) = n_M m(\bgamma)/n(\bgamma)$ satisfies \eqref{nM} for any well defined $m(\bgamma)$, and $n(\bgamma)/n = \pi m( \bgamma) + (1-\pi) u( \bgamma)$ by definition. 

Next, apart from a maximal MEC set, one would need to accept discordant pairs. In the SL setting, one observes the EDF of $\bgamma$ over $M$, giving rise to $\hat{\theta}_k = n_M(1;k)/n_M$, where $n_M(1;k)$ is the number of agreements on the $k$-th key variable over $M$. The perfect MEC set $\hat{M}$ should have these agreement rates. We have then, for $k=1, ..., K$, 
\begin{equation} \label{thetak}
\hat{\theta}_k = \frac{1}{|\hat{M}|} \sum_{(a,b)\in \hat{M}} \mathbb{I}(\gamma_{ab,k} =1) \qquad\text{for}\quad |\hat{M}| = n_M.
\end{equation}

Thus, no matter how one models $m(\bgamma)$, the perfect MEC set should satisfy jointly the $K+1$ equations defined by \eqref{nM} and \eqref{thetak}, given the knowledge of $r(\bgamma)$.

\section{MEC for unsupervised record linkage}
\label{sec:MECunsup}

Let $\bz$ be the $K$-vector of key variables, which may be imperfect for two reasons: it is not rich enough if the true $\bz$-values are not unique for each distinct entity underlying the two files to be linked, or it may be subjected to errors if the observed $\bz$ is not equal to its true value. Let $A$ contain only the distinct $\bz$-vectors from the first file, after removing any other record that has a duplicated $\bz$-vector to some record that is retained in $A$. In other words, if the first file initially contains two or more records with exactly the same value of the combined key, then only one of them will be retained in $A$ for record linkage to the second file. Similarly let $B$ contain the unique records from the second file. The reason for \emph{separate deduplication of keys} is that no comparisons between the two files can distinguish among the duplicated $\bz$ in either file, which is an issue to be resolved otherwise.    

Given $A$ and $B$ preprocessed as above, the maximal MEC set $M_1$ only consists of the record pairs with the perfect agreement of all the key variables. For probabilistic linkage beyond $M_1$, one can follow the same scheme of MEC in the supervised setting, as long as one is able to obtain an estimate of the probability ratio, given which one can form the MEC set of any chosen size. Nevertheless, to estimate the associated FLR \eqref{psi} and MMR \eqref{tau}, an estimate of $n_M$ is also needed.

\subsection{Algorithm of unsupervised MEC} 
\label{algorithm}

The idea now is to apply \eqref{nM} and \eqref{thetak} jointly. Since setting $\hat{n}_M = |M_1|$ and $\hat{\theta}_k \equiv 1$ associated with the maximal MEC set satisfies \eqref{nM} and \eqref{thetak} automatically, probabilistic linkage requires one to assume $n_M > |M_1|$ and $\theta_k < 1$ for at least some of $k= 1, ..., K$. Moreover, unless there is external information that dictates it otherwise, one can only assume common support $\mathcal{S}(M) = \mathcal{S}(U)$ in the unsupervised setting. Let
\begin{equation} \label{r}
r(\bgamma) = m(\bgamma; \btheta) / u(\bgamma; \boldsymbol{\xi})
\end{equation} 
where the probability of observing $\bgamma$ is $m(\bgamma; \btheta)$ by \eqref{m-model} given that a randomly selected record pair from $\Omega$ belongs to $M$, and $u(\bgamma; \boldsymbol{\xi})$ otherwise, similarly given by \eqref{m-model} with parameters $\xi_k$ instead of $\theta_k$. An iterative algorithm of unsupervised MEC is given below. 
\begin{itemize}
\item[I.] Set $\btheta^{(0)} = (\theta_1^{(0)}, \ldots, \theta_K^{(0)})$ and $n_M^{(0)} = |M_1|$, where $M_1$ is the maximal MEC set. 

\item[II.] {For the $t$-th iteration, let ${g}_{ab}^{(t)} = 1$ if $(a,b) \in M^{(t)}$, and 0 otherwise.}   
\begin{itemize}
\item[i.] {Update $u(\bgamma; \bxi^{(t)})$ by  using (\ref{xi-approx}), which is discussed below, given $\bg^{(t)}=\{g_{ab}^{(t)}: (a,b) \in \Omega\}$, and calculate}
\begin{eqnarray} 
\theta_k^{(t)} &=& \frac{1}{|M^{(t)}|} \sum_{(a,b) \in \Omega} {g}_{ab}^{(t)} \mathbb{I}(\gamma_{ab,k}=1), \label{thetak-MEC} 
\end{eqnarray}
{which maximize $D_M$ in (\ref{ent}) for given $u(\bgamma; \bxi^{(t)}), M^{(t)} = \{(a,b)\in \Omega : {g}_{ab}^{(t)}=1\}$ and $|M^{(t)}|=\sum_{(a,b)\in \Omega} {g}_{ab}^{(t)}$. Once  $\btheta^{(t)}$ and $\bxi^{(t)}$ are obtained, we can update $n_M^{(t)}  = \sum_{\bgamma} n(\bgamma) \hat{g}^{(t)}(\bgamma)$, where}
\begin{align*}
& \hat{g}^{(t)}(\bgamma) \equiv \hat{g}(\bgamma; \btheta^{(t)}, \bxi^{(t)})= \min \Big\{ \frac{|M^{(t)}| r^{(t)}(\bgamma)}{|M^{(t)}| \big( r^{(t)}(\bgamma) -1\big) + n},~ 1\Big\} \\
& r^{(t)}(\bgamma) \equiv r(\bgamma; \btheta^{(t)}, \bxi^{(t)}) = \frac{m(\bgamma; \btheta^{(t)})}{u(\bgamma; \bxi^{(t)})}.
\end{align*}

\item[ii.] {For given $\btheta^{(t)}, \bxi^{(t)}$ and $n_M^{(t)}$, we find the MEC set $M^{(t+1)}=\{(a,b)\in \Omega: g_{ab}^{(t+1)}=1\}$ such that $|M^{(t+1)}|= n_M^{(t)}$ by deduplication in the descending order of $r^{(t)}(\bgamma_{ab})$ over $\Omega$. It maximizes the entropy denoted by $Q^{(t)}(\bg)$}:
\begin{eqnarray}
Q^{(t)}(\bg ) \equiv Q(\bg \mid \bpsi^{(t)}) &=& \frac{1}{n_M^{(t)}} \sum_{(a,b) \in \Omega} {g}_{ab} \log r^{(t)}(\bgamma_{ab}), \label{qfct}
\end{eqnarray}
with respect to $\bg$.
\item[III.] Iterate until $n_M^{(t)} = n_M^{(t+1)}$ or $\|\btheta^{(t)} - \btheta^{(t+1)}\| < \epsilon$, where $\epsilon$ is a small positive value. 
\end{itemize}
\end{itemize}

{A theoretical convergence property of the proposed algorithm and its proof are presented in Appendix A.}

Notice that, insofar as $\Omega = M\cup U$ is highly imbalanced, where the prevalence of $g_{ab}= 1$ is very close to 0, one could simply ignore the contributions from $M$ and use 
\begin{equation} \label{xi-approx}
\hat{{\xi}}_k = \frac{1}{n} \sum_{(a,b)\in \Omega} \mathbb{I}(\gamma_{ab,k} = 1)
\end{equation}
under the model \eqref{m-model} of $u(\bgamma ; \boldsymbol{\xi})$, in which case there is no updating of $u(\bgamma; \boldsymbol{\xi}^{(t)})$. 
Other possibilities of estimating $u(\bgamma; \boldsymbol{\xi})$ will be discussed in {Section \ref{sec:discussion-MLE}}. 

Table \ref{tab:MEC} provides an overview of MEC for record linkage in the supervised or unsupervised setting. In the supervised setting, one observes $\bgamma$ for the matched record pairs in $M$, so that the probability $m(\bgamma)$ can be estimated from them directly. Whereas, for MEC in the unsupervised setting, one cannot separate the estimation of $m(\bgamma)$ and $n_M$.

\begin{table}[ht]
\caption{MEC for record linkage in supervised or unsupervised setting}
\centering
\begin{tabular}{l | c | c}  \hline
& Supervised & Unsupervised \\ \hline
$\Omega = M \cup U$ & Observed & Unobserved \\ \hline
\multirow{2}{*}{Probability ratio} & $r_f(\bgamma)$ generally applicable & $r(\bgamma)$ generally \\ 
& $r(\bgamma)$ given $\mathcal{S}(M) \subseteq \mathcal{S}(U)$ & assuming $\mathcal{S}(M) =\mathcal{S}(U)$ \\ \hline
\multirow{2}{*}{Model of $\bgamma$} & \multicolumn{2}{c}{Multinomial if only discrete comparison scores} \\
& \multicolumn{2}{c}{Directly or via key variables and measurement errors} \\ \hline
\multirow{2}{*}{MEC set} & \multicolumn{2}{c}{Guided by FLR and MMR} \\
& \multicolumn{2}{c}{Require estimate of $n_M$ in addition} \\ \hline
\multirow{2}{*}{Estimation} & $m(\bgamma;\btheta)$ from $\bgamma_M$ in $\Omega$ & $m(\bgamma;\btheta)$ and $n_M$ \\
& $n_M$ by \eqref{nM} outside $\Omega$ & jointly by \eqref{nM} and \eqref{thetak} \\ \hline
\end{tabular} \label{tab:MEC}
\end{table}

\subsection{Error rates} \label{MEC-FLR}

MEC for record linkage should generally be guided by the error rates, FLR and MMR, without being restricted to the estimate of $n_M$. 

Note that $\{ \hat{g}_{ab} : (a,b)\in \hat{M}\}$ of any MEC set $\hat{M}$ are among the largest ones over $\Omega$, because MEC follows the descending order of $\hat{r}_{ab}$, except for necessary deduplication when there are multiple pairs involving a given record. To exercise greater control of the FLR, let $\psi$ be the target FLR, and consider the following bisection procedure.
\begin{itemize}
\item[i.] Choose a threshold value $c_{\psi}$ and form the corresponding MEC set $\hat{M}(c_{\psi})$, where $\hat{r}_{ab} \geq c_{\psi}$ for any $(a,b) \in \hat{M}(c_{\psi})$. 

\item[ii.] Calculate the estimated FLR of the resulting MEC set $\hat{M}$ as
\begin{equation}\label{FLR}
\hat{\psi} = \frac{1}{|\hat{M}|} \sum_{(a,b)\in \hat{M}} (1- \hat{g}_{ab}).
\end{equation}
If $\hat{\psi} > \psi$, then increase $c_{\psi}$; if $\hat{\psi} < \psi$, then reduce $c_{\psi}$. 
\end{itemize} 
Iteration between the two steps would eventually lead to a value of $c_{\psi}$ that makes $\hat{\psi}$ as close as possible to $\psi$, for the given probability ratio $\hat{r}(\bgamma)$.

The final MEC set $\hat{M}$ can be chosen in light of the corresponding FLR estimate $\hat{\psi}$. It is also possible to take into consideration the estimated MMR given by
\begin{equation} \label{MMR}
\hat{\tau} = 1 - \sum_{(a,b)\in \hat{M}} \hat{g}_{ab} / \hat{n}_M
\end{equation}
where $\hat{n}_M$ is given by unsupervised MEC algorithm. Note that if $|\hat{M}| = \hat{n}_M$, then we shall have $\hat{\psi} = \hat{\tau}$; but not if $\hat{M}$ is guided by a given target value of FLR or MMR.

{In Section 6.2, we investigate the performance of the MEC sets guided by the error rates through simulations.}


\section{Discussion} 
\label{sec:discussion}

Below we discuss and compare two other approaches in the unsupervised setting, including the ways by which some of their elements can be incorporated into the MEC approach. Other less practical approaches are discussed in Appendix C.

\subsection{The classical approach}

Recall Problems I and II of the classical approach mentioned in Section \ref{sec:problems}. 

From a practical point of view, Problem I can be dealt with by any deduplication method of the set $M^*$ of classified records pairs, where $\hat{r}(\bgamma_{ab})$ is above a threshold value for all $(a,b) \in M^*$. {As ``an advance over previous ad hoc assignment methods'', \cite{jaro1989advances} chooses the linked set $\hat{M}^* \subseteq M^*$, which maximises the sum of $\log \hat{r}(\bgamma_{ab})$ subject to the constraint of one-one link. Since $\hat{g}_{ab}$ is a monotonic function of $\hat{r}(\bgamma_{ab})$, this amounts to choose $\hat{M}^*$ which maximises the expected number of matches in it, denoted by
\[
n_M^* = \sum_{(a,b)\in \hat{M}^*} \hat{g}_{ab}
\]
But $n_M^*$ is still not connected to the probabilities of false links and non-links defined by \eqref{FS}. As illustrated below, neither does it directly control the errors of the linked $\hat{M}^*$.} 

{Consider linking two files with 100 records each. Suppose Jaro's assignment method yields $|\hat{M}^*| =100$ on one occasion, where 80 links have $\hat{g}_{ab} \approx 1$ and 20 links have $\hat{g}_{ab} \approx 0.75$, such that $n_M^* \approx 95$. Suppose it yields 90 links with $\hat{g}_{ab} \approx 1$ and 10 links with $\hat{g}_{ab} \approx 0.5$ on another occasion, where $n_M^* \approx 95$. Clearly, $n_M^*$ does not directly control the linkage errors in $\hat{M}^*$. Moreover, there is no compelling reason to accept 100 links on both these occasions, simply because 100 one-one links are possible.}

In forming the MEC set one deals with Problem I directly, based on the concept of maximum entropy that has relevance in many areas of scientific investigation. The implementation is simple and fast for large datasets. The estimated error rates FLR \eqref{FLR} and MMR in \eqref{MMR} are directly defined for a given MEC set. 


Problem II concerns the parameter estimation. As explained earlier, applying the EM algorithm based on the objective function (\ref{jaro}) proposed by \cite{winkler1988} and \cite{jaro1989advances} is \emph{not} a valid approach of maximum likelihood estimation (MLE). 
One may easily compare this WJ-procedure to that given in Section \ref{algorithm}, where both adopt the same model \eqref{m-model} and the same estimator of $u(\bgamma;\boldsymbol{\xi})$ via $\hat{\xi}_k$ given by \eqref{xi-approx}. It is then clear that the same formula is used for updating $n_M^{(t)}$ at each iteration, but a different formula is used for
\begin{equation} \label{thetak-jaro}
\theta_k^{(t)} = \frac{1}{{n}_M^{(t)}} \sum_{(a,b) \in \Omega} \hat{g}_{ab}^{(t)} \gamma_{ab,k}  
\end{equation}
where the numerator is derived from \emph{all} the pairs in $\Omega$, whereas $\theta_k^{(t)}$ given by \eqref{thetak-MEC} uses only the pairs in the MEC set $M^{(t)}$. Notice that the two differ only in the unsupervised setting, but they would become the same in the supervised setting, where one can use the observed binary $g_{ab}$ instead of the estimated fractional $\hat{g}_{ab}$. 

Thus, one may incorporate the WJ-procedure as a variation of the unsupervised MEC algorithm, where the formulae \eqref{thetak-jaro} and \eqref{xi-approx} are chosen specifically. This is the reason why it can give reasonable parameter estimates in many situations, despite its misconception as the MLE. Simulations will be used later to compare empirically the two formulae \eqref{thetak-MEC} and \eqref{thetak-jaro} for $\theta_k^{(t)}$.

\subsection{An approach of MLE}
\label{sec:discussion-MLE}


{Below we derive another estimator of $\xi_k$ by the ML approach, which can be incorporated into the proposed MEC algorithm, instead of (\ref{xi-approx}).} This requires a model of the key variables, which explicates the assumptions of key-variable errors. Let $z_k$ be the $k$-th key variable which takes value $1, ..., D_k$. \cite{copas1990record} envisage a non-informative hit-miss generation process, where the observed $z_k$ can take the true value despite the perturbation. \cite{copas1990record} demonstrate that the hit-miss model is plausible in the SL {(Supervised Learning)} setting based on labelled datasets.




We adapt the hit-miss model to the unsupervised setting as follows. First, for any $(a,b) \in M$, let $\alpha_k = \mbox{Pr}(e_{ab,k} = 1)$, where $e_{ab,k} =1$ if the associated pair of key variables are subjected to \emph{any form of perturbation} that could potentially cause disagreement of the $k$-th key variable, and $e_{ab,k} =0$ otherwise. Let
\begin{align*}
\theta_k = (1 - \alpha_k) + \alpha_k \sum_{d=1}^{D_k} m_{kd}^2 = 1 - \alpha_k (1 - \sum_{d=1}^{D_k} m_{kd}^2)
\end{align*}
where we assume that $\alpha_k$ must be positive for some $k=1, ..., K$, and 
\[
m_{kd} = \mbox{Pr}(z_{ik} = d | g_{ab} = 1, e_{ab,k} =1) = \mbox{Pr}(z_{ik} = d | g_{ab} = 1, e_{ab,k} =0) 
\] 
for $i=a$ or $b$. Next, for any record $i$ in either $A$ or $B$, let $\delta_i = 1$ if it has a match in the other file and $\delta_i = 0$ otherwise. Given $\delta_i = 0$, with or without perturbation, let
$\mbox{Pr}(z_{ik} = d | \delta_i = 0) = u_{kd}$.
We have $\beta_{kd} := m_{kd}\equiv u_{kd}$ if $\delta_i$ is \emph{non-informative}. 
A slightly more relaxed assumption is that $\delta_i$ is only non-informative in one of the two files. To be more resilient against its potential failure, one can assume $m_{kd}$ to hold for all the records in the \emph{smaller} file, and allow $u_{kd}$ to differ for the records with $\delta_i = 0$ in the \emph{larger} file. Suppose $n_A < n_B$. Let 
\[
p = \mbox{Pr}(\delta_b =1) = E(n_M)/n_B = n_A \pi
\]
be the probability that a record in $B$ has a match in $A$. 
One may assume $\bz_A = \{ \bz_a : a\in A\}$ to be independent over $A$, giving 
\[
\ell_A = \sum_{a\in A} \sum_{k=1}^K \log m_{ak} 
\]
where $m_{ak} = \sum_{d=1}^{D_k} m_{kd} \mathbb{I}(z_{ak} = d)$. 
The complete-data log-likelihood based on $(\delta_B, \bz_B)$ is 
\begin{equation} \label{ell}
\ell_B = \sum_{b\in B} \delta_b \log \Big( p \prod_{k=1}^K m_{bk} \Big) 
+ \sum_{b\in B} (1 -\delta_b) \log \Big( (1-p) \prod_{k=1}^K u_{bk} \Big)
\end{equation}
where $m_{bk} = \sum_{d=1}^{D_k} m_{kd} \mathbb{I}(z_{bk} = d)$ and $u_{bk} = \sum_{d=1}^{D_k} u_{kd} \mathbb{I}(z_{bk} = d)$,   based on an assumption of independent $(\delta_b, \bz_b)$ across the entities in $B$. 

Under separate modelling of $\bz_A$ and $(\bz_B, \delta_B)$, let $\hat{m}_{kd}$ be the MLE based on $\ell_A$, given which an EM-algorithm for estimating $p$ and $u_{kd}$ follows from \eqref{ell} by treating $\delta_B$ as the missing data. However, the estimation is feasible only if $\{ u_{kd}\}$ and $\{ m_{kd}\}$ are not exactly the same; whereas the MLE of $n_M$ has a large variance, when $\{ m_{kd}\}$ and $\{ u_{kd}\}$ are close to each other, even if they are not exactly equal. 


Meanwhile, the closeness between $\{ m_{kd}\}$ and $\{ u_{kd}\}$ does not affect the MEC approach, where $\hat{n}_M$ is obtained from solving \eqref{nM} given $\hat{r}(\bgamma) = \hat{m}(\bgamma)/\hat{u}(\bgamma)$, where $\hat{u}(\bgamma)$ is indeed most reliably estimated when $\{ m_{kd}\} = \{ u_{kd}\}$. Moreover, one can incorporate a \emph{profile EM-algorithm}, based on \eqref{ell} given $n_M^{(t)}$, to update $u(\bgamma; \boldsymbol{\xi}^{(t)})$ in the unsupervised MEC algorithm of Section \ref{algorithm}. At the $t$-th iteration, where $t\geq 1$, given $p^{(t)} = n_M^{(t)}/\max(n_A, n_B)$ and $\hat{m}_{kd}$ estimated from the smaller file $A$, obtain $u_{kd}^{(t)}$ by 
\begin{equation} \label{xi-profile}
{\xi}_k^{(t)} = \Big( (1-p^{(t)}) \sum_{d=1}^{D_k} u_{kd}^{(t)} \hat{m}_{kd} + p^{(t)}(1-\frac{1}{n_A}) \sum_{d=1}^{D_k} \hat{m}_{kd}^2 \Big)
/ \big( 1-p^{(t)}/n_A \big). 
\end{equation}


\section{Simulation Study }
\label{sec:simul}

\subsection{Set-up}

To explore the practical feasibility of the unsupervised MEC algorithm for record linkage, we conduct a simulation study based on the data sets listed in Table \ref{sec2:tb1}, which are disseminated by ESSnet-DI \citep{mcleod2011simulated} and freely available online. 
Each record in a data set has associated synthetic key variables, which may be distorted by missing values and typos when they are created, in ways that imitate real-life errors {\citep{mcleod2011simulated}}. 

\begin{table}[ht]
\centering
\caption{Data set description (size in parentheses)} 
\label{sec2:tb1}
\begin{tabular}{c|c} \hline
{Data set}  &  Description \\ \hline
{Census} & A fictional data set to represent some observations \\
$(25,343)$ & from a decennial Census\\ \hline
{CIS} & Fictional observations from Customer Information System, \\
$(24,613)$ & combined administrative data from the tax and benefit systems\\ \hline
{PRD} & Fictional observations from Patient Register Data\\
$(24,750)$ &  of the National Health Service \\ \hline
\end{tabular}
\end{table}

We consider the linkage keys forename, surname, sex, and date of birth (DOB). To model the key variables, we divide DOB into 3 key variables (Day, Month, Year). For text variables such as forename and surname, we divide them into 4 key variables by using the Soundex coding algorithm \citep[][p. 290]{copas1990record}, which reduces a name to a code consisting of the leading letter followed by three digits, e.g. Copas $\equiv$ C120, Hilton $\equiv$ H435. The twelve key variables for record linkage are presented in Table \ref{sec2:tb2}. 

\begin{table}[ht]
\centering
\caption{Twelve key variables available in the three data sets.} 
\label{sec2:tb2}
\begin{tabular}{cc|c|c} \hline
\multicolumn{2}{c|}{Variable}  &  Description  & No. of Categories\\ \hline
\multirow{4}{*}{PERNAME1} & 1 & First letter of forename & 26\\
& 2 & First digit of Soundex code of forename & 7 \\
& 3 & Second digit of Soundex code of forename & 7 \\
& 4 & Third digit of Soundex code of forename & 7 \\ \hline
\multirow{4}{*}{PERNAME2}& {1} &  First letter of surname & 26\\
& {2} &First digit of Soundex code of surname & 7 \\
& {3} & Second digit of Soundex code of surname & 7 \\
& {4} & Third digit of Soundex code of surname & 7 \\ \hline
\multicolumn{2}{c|}{SEX} & Male / Female & 2\\ \hline
\multirow{3}{*}{DOB}& {DAY} & Day of birth & 31\\
& {MON} & Month of birth & 12\\
& {YEAR} & Year of birth  (1910 $\sim$ 2012) & 103\\ \hline
\end{tabular}
\end{table}

We set up two scenarios to generate linkage files. We use the unique identification variable (PERSON-ID) for sampling, which are available in all the three data sets. We sample $n_A = 500$ and $n_B=1000$ individuals from PRD and CIS, respectively. Let $p_A$ be the proportion of records in the smaller file (PRD) that are also selected in the larger file (CIS), by which we can vary the degree of overlap, i.e. the set of matched individuals $AB$, between $A$ and $B$. We use $p_A = 0.8, 0.5$ or $0.3$ under either scenario. 


\paragraph{\textnormal{\em Scenario-I (Non-informative)}}
\begin{itemize}
\item Sample $n_0= n_B/p_A$ individuals randomly from Census.
\item Sample $n_A$ randomly from these $n_0$ as the individuals of PRD, denoted by $A$.
\item Sample $n_B$ randomly from these $n_0$ as the individuals of CIS, denoted by $B$.
\end{itemize}
Under this scenario both $\delta_a$ and $\delta_b$ are non-informative for the key-variable distribution. For any given $p_A$, we have $E(n_M) = n_A p_A$ and $\pi = E(n_M)/n_0$, where $n_M$ is the random number of matched individuals between the simulated files $A$ and $B$.

\paragraph{\textnormal{\em Scenario-II (Informative)}}
\begin{itemize}
\item Sample $n_A$ randomly from Census $\cap$ PRD $\cap$ CIS, denoted by $A$ from PRD.
\item Sample $n_M = n_A p_A$ randomly from $A$ as the matched individuals, denoted by $AB$.
\item Sample $n_B - n_M$ randomly from CIS $\setminus$ A having $\mathrm{SEX}=F$,    $\mathrm{YEAR} \le 1970$, and odd $\mathrm{MON}$, denoted by $B_0$. Let $B = AB \cup B_0$ be the sampled individuals of CIS.
\end{itemize}
Under this scenario the key-variable distribution is the same in $A$, whether or not $\delta_a =1$, but it is different for the records $b\in B_0$, or $\delta_b = 0$. Hence, scenario-II is informative. For any given $p_A$, we have fixed $n_M = n_A p_A$ and $\pi = p_A/n_B$.

\subsection{Results: Estimation}

For the unsupervised MEC algorithm given in Section \ref{algorithm}, one can adopt \eqref{thetak-MEC} or \eqref{thetak-jaro} for updating $\theta_k^{(t)}$. Moreover, one can use \eqref{xi-approx} for $\hat{\xi}_k$ directly, or \eqref{xi-profile} for updating $\xi_k^{(t)}$ iteratively. In particular, choosing \eqref{thetak-jaro} and \eqref{xi-approx} effectively incorporates {the procedure of \cite{winkler1988} and \cite{jaro1989advances}} for parameter estimation. Note that the MEC approach still differs to that of \cite{jaro1989advances}, with respect to the formation of the linked set $\hat{M}$.

Table \ref{simul:tb1} compares the performance of the unsupervised MEC algorithm, using different formulae for $\theta_k^{(t)}$ and $\xi_ k^{(t)}$, where the size of $\hat{M}$ is equal to the corresponding estimate $\hat{n}_M$. In addition, we include $\hat{\theta}_k = n_M(1;k)/n_M$ estimated directly from the matched pairs in $M$, {as if $M$ were available for supervised learning}, together with \eqref{xi-approx} for $\hat{\xi}_k$. The true parameters and error rates are given in addition to their estimates.

\FloatBarrier
\begin{table}[ht]
\centering
\caption{Parameters and averages of their estimates, averages of error rates and their estimates, over 200 simulations. Median of estimate of $n_M$ given as $\tilde{n}_M$.} 
\label{simul:tb1}
\begin{tabular}{cc|cc|ccccccc} \hline
\multicolumn{11}{c}{Scenario I}\\ \hline
\multicolumn{2}{c|}{Parameter} & \multicolumn{2}{c|}{Formulae} & \multicolumn{7}{c}{Estimation} \\ \cline{1-2}\cline{3-4}\cline{5-11}
$\pi$ & $E(n_M)$ & $\theta_k^{(t)}$ & $\xi_k^{(t)}$ & $\hat{\pi}$ & $\hat{n}_M$ & $\tilde{n}_M$ & $\mbox{FLR}$ & $\mbox{MMR}$ & $\widehat{\mbox{FLR}}$ & $\widehat{\mbox{MMR}}$\\ \hline
\multirow{4}{*}{.0008} & \multirow{4}{*}{400}  & $\hat{\theta}_k$ &  \eqref{xi-approx} 
& .00080 & 400.0 & 397 & .0264 & .0266 & .0357 & .0357 \\ 
& & \eqref{thetak-MEC} & \eqref{xi-profile} & .00082 & 407.9 & 405 & .0425 & .0257 & .0509 & .0509 \\
& & \eqref{thetak-MEC} &\eqref{xi-approx} &  .00083 & 414.7 & 407& .0549 & .0244 & .0620 & .0620 \\
& & \eqref{thetak-jaro} & \eqref{xi-approx} & .00081 & 406.0 & 405 &.0399  & .0269 & .0503 & .0503 \\ \hline
\multirow{4}{*}{.0005} & \multirow{4}{*}{250} & $\hat{\theta}_k$ & \eqref{xi-approx}  
& .00050 & 251.6 & 249 & .0340 & .0301 & .0370 & .0370 \\
 & & \eqref{thetak-MEC} & \eqref{xi-profile} & .00052 & 258.3 & 255 & .0559 & .0296 & .0533 & .0533\\
 & & \eqref{thetak-MEC} & \eqref{xi-approx} & .00053 & 266.9 & 256.5 & .0742 & .0277 & .0680 & .0680 \\
&  & \eqref{thetak-jaro} & \eqref{xi-approx} & .00052 & 261.7 & 259 &.0676 & .0305 & .0636 & .0636 \\ \hline
\multirow{4}{*}{.0003} & \multirow{4}{*}{150} & $\hat{\theta}_k$ &  \eqref{xi-approx}  
& .00030 & 152.3 & 151 & .0439 & .0356 & .0381 & .0381 \\
 & & \eqref{thetak-MEC} & \eqref{xi-profile} & .00033 & 165.9 & 156.5 &.0873 & .0244 & .0620 & .0620 \\
& & \eqref{thetak-MEC} &  \eqref{xi-approx} & .00041 & 205.4 & 161 & .1632 & .0308 & .1251 & .1251 \\
&  & \eqref{thetak-jaro} & \eqref{xi-approx} & .00054 & 271.4 & 169 &.3015 & .0785 & .1639 & .1639 \\  \hline 
\multicolumn{10}{c}{Scenario II}\\ \hline
\multicolumn{2}{c|}{Parameter} & \multicolumn{2}{c|}{Formulae} & \multicolumn{7}{c}{Estimation} \\ \cline{1-2}\cline{3-4}\cline{5-11}
$\pi$ & $n_M$ & $\theta_k^{(t)}$ & $\xi_k^{(t)}$ & $\hat{\pi}$ & $\hat{n}_M$ & $\tilde{n}_M$ & $\mbox{FLR}$ & $\mbox{MMR}$ & $\widehat{\mbox{FLR}}$ & $\widehat{\mbox{MMR}}$\\ \hline
\multirow{4}{*}{.0008} & \multirow{4}{*}{400}  & $\hat{\theta}_k$ &  \eqref{xi-approx}  
& .00080 & 398.3 & 400  & .0230 & .0273 & .0326 & .0326 \\ 
& & \eqref{thetak-MEC} &  \eqref{xi-profile} &  .00080 & 401.4 & 401  & .0305 & .0277 & .0403 & .0403 \\
& & \eqref{thetak-MEC} & \eqref{xi-approx} &  .00081 & 405.2 & 404 & .0379 & .0262 & .0467 & .0467 \\
& & \eqref{thetak-jaro} & \eqref{xi-approx} & .00080 & 401.4 & 401 & .0316 & .0286 & .0438 & .0438 \\ \hline
\multirow{4}{*}{.0005} & \multirow{4}{*}{250} & $\hat{\theta}_k$ & \eqref{xi-approx}  
& .00050 & 249.6 & 250 & .0284 & .0302 & .0334 & .0334 \\
 & & \eqref{thetak-MEC} & \eqref{xi-profile} & .00050 & 251.8 & 251 & .0383 & .0320 & .0410 & .0410 \\
 & & \eqref{thetak-MEC} & \eqref{xi-approx} & .00052 & 257.7 & 253 & .0513 & .0295 & .0516 & .0516 \\
&  & \eqref{thetak-jaro} & \eqref{xi-approx} & .00051 & 255.4 & 253.5 & .0510 & .0336 & .0520 & .0520 \\ \hline
\multirow{4}{*}{.0003} & \multirow{4}{*}{150} & $\hat{\theta}_k$ &  \eqref{xi-approx} 
& .00030 & 150.5 & 150 & .0382 & .0355 & .0350 & .0350 \\
 & & \eqref{thetak-MEC} & \eqref{xi-profile} & .00031 & 153.0 & 153 & .0559 & .0377 & .0452 & .0452 \\
& & \eqref{thetak-MEC} & \eqref{xi-approx} & .00032 & 158.5 & 155 & .0708 & .0342 & .0558 & .0558 \\
&  & \eqref{thetak-jaro} & \eqref{xi-approx} & .00038 & 189.3 & 156 & .1414 & .0524 & .0903 & .0903 \\ \hline
\end{tabular}
\end{table}
\FloatBarrier

\begin{figure}
\centering
\includegraphics[width=14cm, height=6cm]{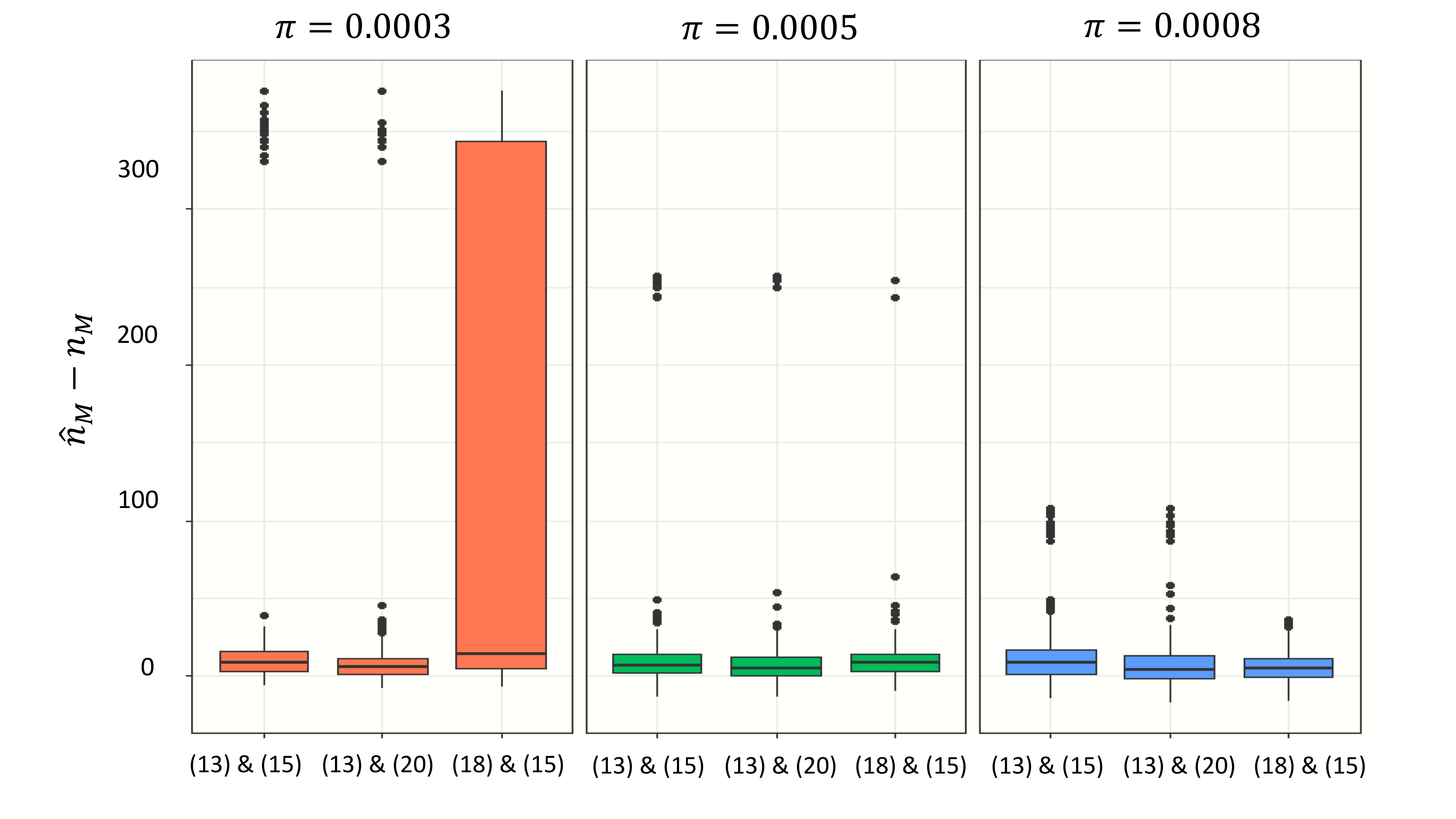}
\caption{Box plots of $\hat{n}_M -n_M$ based on 200 Monte Carlo samples under Scenario I. {\label{fig1}}}
\end{figure}

\begin{figure}
\centering
\includegraphics[width=14cm, height=6cm]{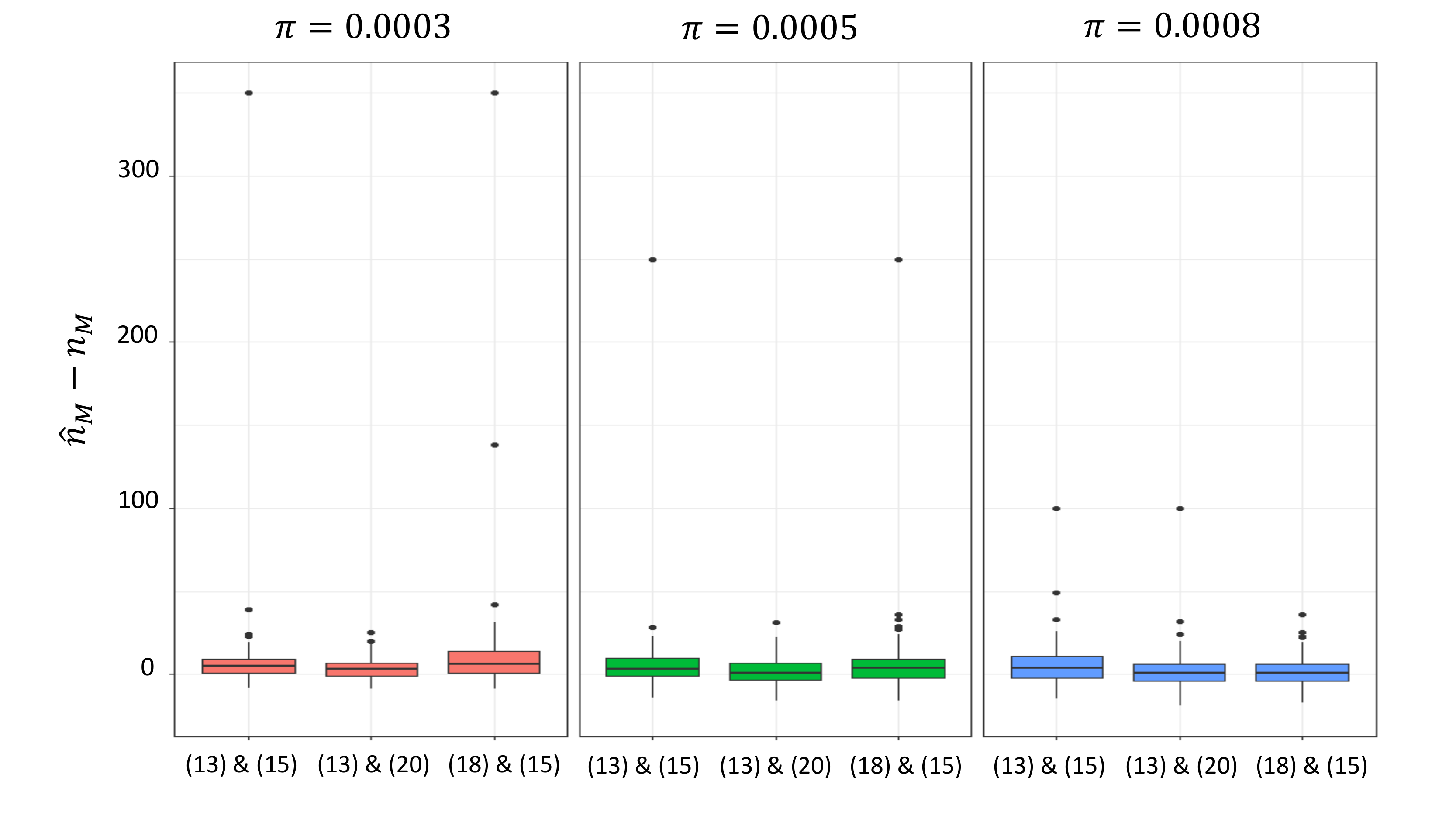}
\caption{Box plots of $\hat{n}_M -n_M$ based on 200 Monte Carlo samples under Scenario II. {\label{fig2}}}
\end{figure}

{As expected, the best results are obtained when the parameter $\theta_k$ is estimated directly from the matched pairs in $M$, i.e.,  $\hat{\theta}_k = n_M(1;k)/n_M$, together with (\ref{xi-approx}) for $\hat{\xi}_k$, despite $\hat{\xi}_k$ by (\ref{xi-approx}) is not exactly unbiased.} Nevertheless, the approximate estimator $\hat{\xi}_k$ can be improved, since the profile-EM estimator given by \eqref{xi-profile} is seen to perform better across all the set-ups, where both are combined with \eqref{thetak-MEC} for $\theta_k^{(t)}$. When it comes to the two formulae of $\theta_k^{(t)}$ by \eqref{thetak-MEC} and \eqref{thetak-jaro}, and the resulting $n_M$-estimators and the error rates FLR and MMR, we notice the followings. 
\begin{itemize}

\item Scenario-I: When the size of the matched set $M$ is relatively large at $p_A = 0.8$ , there are only small differences in terms of the average and median of the two estimators of $n_M$, and the difference is just a couple of false links in terms of the linkage errors.   Figures \ref{fig1} shows that \eqref{thetak-MEC} results in a few larger errors of $\hat{n}_M$ than \eqref{thetak-jaro} over the 200 simulations, {when $p_A =0.8$ or $\pi = 0.0008$.} As the size of the matched set $M$ decreases, the averages and medians of the estimators of $n_M$ resulting from \eqref{thetak-MEC} and \eqref{xi-profile} are closer to the true values than those of the other estimators. Especially when the matched set $M$ is relatively small, where $\pi = 0.0003$, the formula \eqref{thetak-jaro} results in considerably worse estimation of $n_M$ in every respect. While this is partly due to the use of \eqref{xi-approx} instead of \eqref{xi-profile}, most of the difference is down to the choice of $\theta_k^{(t)}$, which can be seen from intermediary comparisons to the results based on \eqref{thetak-MEC} and \eqref{xi-approx}. 

\item Scenario-II: The use of \eqref{thetak-MEC} and \eqref{xi-profile} for the unsupervised MEC algorithm performs better than using the other formulae in terms of both estimation of $n_M$ and error rates across the three sizes of the matched set (Figure \ref{fig2}). Relatively greater improvement is achieved by using \eqref{thetak-MEC} and \eqref{xi-profile} for the smaller matched sets. 
\end{itemize}

The results suggest that the unsupervised MEC algorithm tends to be more affected by the size of the matched set under Scenario-I than Scenario-II. Choosing \eqref{thetak-MEC} and \eqref{xi-profile}, however, seems to yield the most robust estimation of $n_M$ and error rates against the small size of the matched set $M$, regardless the informativeness of key-variable errors. The reason must be the fact that the numerator of $\theta_k^{(t)}$ is calculated in \eqref{thetak-jaro} over all the pairs in $\Omega$ instead of the MEC set $M^{(t)}$, which seems more sensitive when the imbalance between $M$ and $U$ is aggravated, while the sizes of $A$ and $B$ remain fixed.

{We also include the additional results obtained for $p_A = 0.2, 0.15,$ and $0.1$  in Appendix D. The estimate $\hat{n}_M$ (or $\hat{\pi}$) gets worse as $p_A$ (or $\pi$) reduces, which is consistent with the previous findings of others, for example, \citet{enamorado2019using} showed that a greater degree of overlap between data sets leads to better merging results in terms of the error rates as well as the accuracy of their estimates. The problem is also highlighted by \citet{sadinle2017bayesian}. Record linkage in cases of extremely low prevalence of true matches is a problem that needs to be studied more carefully on its own.}

\subsection{Results: MEC set}

Aiming the MEC set $\hat{M}$ at the estimated size $\hat{n}_M$ is generally not a reasonable approach to record linkage. Record linkage should be guided directly by the associated uncertainty, i.e. the error rates FLR and MMR, based on their estimates \eqref{FLR} and \eqref{MMR}, as described in Section \ref{MEC-FLR}. Note that this does require the estimation of $n_M$ in addition to $r(\boldsymbol{\gamma})$.  
 
We have $\widehat{\mbox{FLR}} = \widehat{\mbox{MMR}}$ in Table \ref{simul:tb1}, because $|\hat{M}| = \hat{n}_M$ here. {It can be seen that these follow the true FLR more closely than the MMR, especially when $\hat{n}_M$ is estimated using the formulae \eqref{thetak-MEC} and \eqref{xi-profile}.} This is hardly surprising. Take e.g. the maximal MEC set $M_1$ that consists of the pairs whose key variables agree completely and uniquely. Provided reasonably rich key variables, as the setting here, one can expect the FLR of $M_1$ to be low, such that even a na\"{i}ve estimate $\widehat{\mbox{FLR}} =0$ probably does not err much. {Meanwhile, the true MMR has a much wider range from one application to another, because the difference between $n_M$ and $|M_1|$ is determined by the extent of key-variable errors, such that the estimate of MMR depends more critically on that of $n_M$.} The situation is similar for any MEC set beyond $M_1$, as long as $\hat{g}_{ab}$ remains very high for any $(a,b)\in \hat{M}$.

\begin{table}[ht]
\centering
\caption{Parameters and averages of their estimates, averages of error rates and their estimates, over 200 simulations, {$n = |\Omega| = n_A n_B$}.} 
\label{simul:tb2}
\begin{tabular}{cc|c|ccccccc} \hline
\multicolumn{9}{c}{Scenario I}\\ \hline
\multicolumn{2}{c|}{Parameter} & Target & \multicolumn{7}{c}{Estimation} \\ \cline{1-2}\cline{4-10}
$\pi$ & $E(n_M)$ & {FLR} & $\hat{n}_M$ & {$|\hat{M}|/n$} & {$|\hat{M}|$} & $\mbox{FLR}$ & $\mbox{MMR}$ & $\widehat{\mbox{FLR}}$ & $\widehat{\mbox{MMR}}$ \\ \hline
\multirow{2}{*}{.0008} & \multirow{2}{*}{400}  &  {0.05}
& \multirow{2}{*}{407.9} & .00080 & 401.9  & .0313 & .0280 & .0393 & .0527 \\ 
& & {0.03} & & .00079 & 395.0  & .0196 & .0328 & .0271 & .0568 \\
\hline
\multirow{2}{*}{.0005} & \multirow{2}{*}{250} & {0.05} & \multirow{2}{*}{258.3} & .00050 & 251.9  & .0396 & .0326 & .0385 & .0576 \\
& & {0.03} &  & .00049 & 246.7  & .0246 & .0374 & .0264 & .0650 \\
\hline
\multirow{2}{*}{.0003} & \multirow{2}{*}{150} & {0.05} 
& \multirow{2}{*}{165.9} & .00031 & 153.4 & .0533 & .0403 & .0389 & .0783 \\
 & & 0.03 &  & .00030 &  149.3  &.0355 & .0483 & .0256 & .0905 \\
\hline
\multicolumn{9}{c}{Scenario II}\\ \hline
\multicolumn{2}{c|}{Parameter} & Target & \multicolumn{7}{c}{Estimation} \\ \cline{1-2}\cline{4-10}
$\pi$ & {$n_M$} & {FLR} & $\hat{n}_M$ & {$|\hat{M}|/n$} & {$|\hat{M}|$} & $\mbox{FLR}$ & $\mbox{MMR}$ & $\widehat{\mbox{FLR}}$ & $\widehat{\mbox{MMR}}$ \\ \hline
\multirow{2}{*}{.0008} & \multirow{2}{*}{400}  &  {0.05}
& \multirow{2}{*}{401.4} & .00080  & 397.8  & .0239 & .0294 & .0337 & .0418 \\
& & {0.03} & & .00079  & 393.1  & .0164 & .0334 & .0256 & .0451 \\
\hline
\multirow{2}{*}{.0005} & \multirow{2}{*}{250} & {0.05} & \multirow{2}{*}{251.8} &  .00050  & 248.6  & .0305 & .0361 & .0328 & .0447 \\
& & {0.03} & & .00049  & 245.2  & .0226 & .0416 & .0245 & .0497 \\
\hline
\multirow{2}{*}{.0003} & \multirow{2}{*}{150} & {0.05} 
& \multirow{2}{*}{153.0} & .00030  & 150.1 & .0445 & .0443 & .0333 & .0514 \\
 & & 0.03 & & .00029  & 147.4 &.0322 & .0489 & .0238 & .0588 \\
\hline
\end{tabular}
\end{table}

Table \ref{simul:tb2} shows the performance of the MEC set using the bisection procedure described in Section \ref{MEC-FLR}, across the same set-ups as in Table \ref{simul:tb1}. We use only \eqref{thetak-MEC} for $\theta_k^{(t)}$ and \eqref{xi-profile} for $\xi_k^{(t)}$ to obtain the corresponding $\hat{n}_M$. We let the target FLR be $\psi = 0.05$ or 0.03, where the latter is clearly lower than the true FLR of $\hat{M}$ that is of the size $\hat{n}_M$ (Table \ref{simul:tb1}), especially when the prevalence is relatively low (at $\pi = 0.0003$) under either scenario. The resulting true (FLR, MMR) and their estimates are given in Table \ref{simul:tb2}.

It can be seen that the MEC algorithm guided by the FLR yields the MEC set $\hat{M}$, whose size $|\hat{M}|$ is close to the true $n_M$ across all the set-ups. Indeed, under Scenario-I, the mean of $|\hat{M}|$ is closer to $n_M$ than the mean (or median) of $\hat{n}_M$ over all the simulations, which results directly from parameter estimation,  especially when the match set is relatively small (at $\pi = 0.0003$) and the performance of $\hat{n}_M$ is most sensitive. In other words, the fact that $|\hat{M}|$ differs to the estimate $\hat{n}_M$ is not necessarily a cause of concern for the MEC algorithm guided by targeting the FLR.
 
To estimate the MMR by \eqref{MMR}, one can either use $|\hat{M}|$ as the estimate of $n_M$, or one can use  $\hat{n}_M$ from parameter estimation based on  \eqref{thetak-MEC} and \eqref{xi-profile}. In the former case, one would obtain $\widehat{\mbox{MMR}} = \widehat{\mbox{FLR}}$. While this $\widehat{\mbox{MMR}}$ is not unreasonable in absolute terms since $|\hat{M}|$ is close to $n_M$ here, as can be seen from comparing the mean of $\widehat{\mbox{FLR}}$ with that of the true MMR in Table \ref{simul:tb2}, it has a drawback \emph{a priori}, in that it decreases as the target FLR decreases, although one is likely to miss out on more true matches when more links are excluded from the MEC set $\hat{M}$. Using $\hat{n}_M$ from parameter estimation directly makes sense in this respect, since the true $n_M$ must remain the same, regardless the target FLR. However, the estimator $\widehat{\mbox{MMR}}$ could then become less reliable given relatively low prevalence $\pi$, where $\hat{n}_M$ could be sensitive in such situations.

In short, the estimation of FLR tends to be more reliable than that of MMR, especially if the prevalence $\pi$ is relatively low in its theoretical range $0< \pi \leq \min(n_A, n_B)/n$. The following recommendations for unsupervised record linkage seem warranted.
\begin{itemize}
\item When forming the MEC set $\hat{M}$ according to the uncertainty of linkage, it is more robust to rely on the FLR, estimated by \eqref{FLR}.
\item The estimate of MMR given by \eqref{MMR}, derived from the parameter estimate $\hat{n}_M$ based on \eqref{thetak-MEC} and \eqref{xi-profile} provides an additional uncertainty measure. However, one should be aware that this measure can be sensitive when the prevalence $\pi$ is relatively low.  
\item Between two target values of the FLR, $\psi < \psi'$, more attention can be given to the estimate of additional missing matches in $\hat{M}(\psi)$ compared to $\hat{M}(\psi')$, given by
\[
\sum_{(a,b)\in \hat{M}(\psi')} \hat{g}_{ab} - \sum_{(a,b)\in \hat{M}(\psi)} \hat{g}_{ab} = \sum_{(a,b)\in \hat{M}(\psi') \setminus \hat{M}(\psi)} \hat{g}_{ab} ~.
\]
\end{itemize}

\section{Final remarks}
\label{sec:final}

We have developed an  approach of maximum entropy classification to record linkage. This provides a unified probabilistic record linkage framework both in the supervised and unsupervised settings, where a coherent classification set of links are chosen explicitly with respect to the associated uncertainty. The theoretical formulation overcomes some persistent flaws of the classical approaches. {Furthermore,} the proposed MEC algorithm is fully automatic, unlike the classical approach that generally requires clerical review to resolve the undecided cases. 

An important issue that is worth further research concerns the estimation of relevant parameters {in} the model of key-variable errors that cause problems for record linkage. First, as pointed out earlier, treating record linkage as a classification problem allows one to explore many modern machine learning techniques. A key challenge in this respect is the fact that the different record pairs are not distinct `units', such that any powerful supervised learning technique needs to be adapted to the unsupervised setting, where it is impossible to estimate the relevant parameters based on the true matches and non-matches, including the number of matched entities. Next, the model of the key-variable errors or the comparison scores can be refined. {Once these issues are resolved together, further improvements on the parameter estimation can be hopefully made,} which will benefit both the classification of the set of links and the assessment of the associated uncertainty. 

Another issue that is interesting to explore in practice is the various possible forms of informative key-variable errors, insofar as the model pertaining to the matched entities in one way or another differs to that of the unmatched entities. Suitable variations of the MEC approach may need to be configured in different situations.

\begin{center}
{\large \bf ACKNOWLEDGMENTS}
\end{center}
The authors thank the associate editor and the reviewers for their constructive comments. Dr. Kim is partially supported by NSF grant MMS 1733572.

\begin{center}
{\large\bf APPENDIX}
\end{center}

\renewcommand{\theequation}{A.\arabic{equation}}
\setcounter{equation}{0}

\section*{A. Convergence property of the proposed algorithm}

{{\bf {\textsl{Theorem 1.}}} {\textsl{Let $\bpsi=\{\bg, \btheta, \bxi\}$ denote the model parameters. The entropy of a classification set of $M$ can be re-written as}}
	\begin{eqnarray*}
		D_M \equiv D(\bpsi) = \frac{1}{n_M(\bpsi)} \sum_{(a,b) \in \Omega} g_{ab} \log r(\bgamma_{ab}; \btheta(\bpsi) , \bxi(\bpsi)), 
	\end{eqnarray*}
	\textsl{and $n_M (\bpsi) = \sum_{\bgamma \in S} n(\bgamma) \hat{g}(\bgamma; \bpsi)$ satisfying the fixed-point equation (9). Let $\bpsi^{(t)} = \{\bg^{(t)}, \btheta^{(t)}, \bxi^{(t)}\}$ denote the current values obtained from the $t$-iteration of the proposed algorithm. {If $Q(\bg^{(t+1)} \mid \bpsi^{(t)}) \ge Q(\bg^{(t)} \mid \bpsi^{(t)})$ holds, where $Q(\bg \mid \bpsi^{(t)})$ is defined as in (14), then $D(\bpsi^{(t+1)}) \ge D(\bpsi^{(t)})$.}}}

{Note that Theorem 1 implies $\{D^{(t)}\}$ is monotone increasing, and it is bounded if $S(M^{(t)}) \subset S(U^{(t)})$. Thus, $\{D(\bpsi^{(t)})\}$ converges to some value, $D^\ast$.}\\

{\bf Proof.} Let 
$$Q(\bg \mid \bpsi') = \frac{1}{n_M(\bpsi')} \sum_{(a,b) \in \Omega} g_{ab} \log r(\gamma_{ab}; \bpsi').$$
\begin{eqnarray*}
	&~& D(\bpsi^{(t+1)}) - D(\bpsi^{(t)}) \\
	&=& D(\bpsi^{(t+1)}) - Q(\bg^{(t+1)} \mid \bpsi^{(t)}) + Q(\bg^{(t+1)} \mid \bpsi^{(t)}) - D(\bpsi^{(t)}) \\
	& \ge & D(\bpsi^{(t+1)} - Q(\bg^{(t+1)} \mid \bpsi^{(t)}) \\
	&=& \frac{1}{n_M^{(t+1)}} \sum_{(a,b) \in \Omega} {g}_{ab}^{(t+1)} \left(\log \frac{m^{(t+1)}(\bgamma_{ab})}{m^{(t)}(\bgamma_{ab})}+\log \frac{u^{(t)}(\bgamma_{ab})}{u^{(t+1)}(\bgamma_{ab})}\right)  + \\
	&~& \left( 1 - \frac{n_M^{(t+1)}}{n_M^{(t)}} \right) \frac{1}{n_M^{(t+1)}} \sum_{(a,b) \in \Omega} g_{ab}^{(t+1)} \log r^{(t)}(\bgamma_{ab}) \\
	&\ge& - \log \left\{ \frac{1}{n_M^{(t+1)}} \sum_{(a,b) \in \Omega} g_{ab}^{(t+1)} \frac{m^{(t)}(\bgamma_{ab})}{m^{(t+1)}(\bgamma_{ab})} \right\} -\log \left\{ \frac{1}{n_M^{(t+1)}} \sum_{(a,b) \in \Omega} g_{ab}^{(t+1)} \frac{u^{(t+1)}(\bgamma_{ab})}{u^{(t)}(\bgamma_{ab})} \right\} -\\ 
	&~& \left|1-\frac{n_M^{(t+1)}}{n_M^{(t)}}\right| \frac{1}{n_M^{(t+1)}} \log \left\{ \sum_{(a,b)\in \Omega} g_{ab}^{(t+1)} \frac{m^{(t)}(\bgamma_{ab})}{u^{(t)}(\bgamma_{ab})}\right\}\\
	& \ge & - \log \left\{ \frac{1}{n_M^{(t+1)}} \sum_{(a,b) \in \Omega} g_{ab}^{(t+1)} m^{(t)}(\bgamma_{ab}) \right\} - \log \left\{ \frac{1}{n_M^{(t+1)}} \sum_{(a,b) \in \Omega} g_{ab}^{(t+1)} u^{(t+1)}(\bgamma_{ab}) \right\} - \\
	&~& \left|1-\frac{n_M^{(t+1)}}{n_M^{(t)}}\right|  \log \left\{\frac{1}{n_M^{(t+1)}} \sum_{(a,b)\in \Omega} g_{ab}^{(t+1)} {m^{(t)}(\bgamma_{ab})}\right\},\\
	&\ge& 0. 
\end{eqnarray*}
The first inequality holds because we have $Q(\bg^{(t+1)} \mid \bpsi^{(t)}) \ge D(\bpsi^{(t)})$ by the construction of the $\bg^{(t+1)}$. The second inequality holds by applying the Jensen's inequality and the condition that $\sum_{(a,b) \in \Omega} g_{ab}^{(t+1)} \log r^{(t)}(\gamma_{ab}) > 0$, Third inequality follows from by multiplying the first term by  $m^{(t+1)}(\bgamma)$ and the second and third terms by $u^{(t)}(\bgamma)$, respectively. Since $\sum_{(a,b) \in \Omega} g_{ab}^{(t+1)} m^{(t)}(\bgamma) = \sum_{\bgamma \in S}n_{M^{(t+1)}}(\bgamma) m^{(t)}(\bgamma) \le n_M^{(t+1)} \sum_{\bgamma \in S} m^{(t)}(\bgamma) = n_M^{(t+1)}$, the last inequality holds.  

\renewcommand{\theequation}{B.\arabic{equation}}
\setcounter{equation}{0}

\section*{B. Special cases of MEC sets for record linkage}

Consider some special cases, given binary $\mathcal{S}$ where $|\mathcal{S}|=2$. The Bernoulli model of $m(\gamma)$ has probability $m(1) = \theta = 1- m(0)$ for $\gamma \in \mathcal{S}(M)$, including $\theta = 1$ or 0 if $|\mathcal{S}(M)| = 1$. Under the Bernoulli model of $f(\gamma)$ in addition, the normalisation constraint is always satisfied by $\hat{\theta}$ obtained from maximising $D_f$ defined over $M$. Every pair $(a,b)$ in a maximal MEC set $M_1$ must have $\gamma_{ab} = 1$. The point we would like to demonstrate is that generally probabilistic record linkage may be guided by the error rates, FLR and MMR, without being restricted to the knowledge (or estimate) of $n_M =|M|$.

\paragraph{\textnormal{\emph{Case-0}}} Suppose one has perfect key variables, such that $\gamma_{ab} = 1$ iff $(a,b)\in M$. Notice that $Q$ by (4) of the main manuscript is not applicable here since $\mathcal{S}(U) \cap \mathcal{S}(M) =\emptyset$. Using $Q_f$ by (3), we have $D_f = n_M \log \theta$, which is maximised at $\hat{\theta} =1$, as well as $S(M) = \{ 1\}$ such that $\hat{\theta} = 1$ follows directly from the normalisation constraint in this case. By (9), we obtain $\hat{g}(1) = 1$ since $r(1) = \infty$, and $\hat{g}(0) =0$ since $r(0) = 0$, so that the estimate of $n_M$ by (8) is $n(1) = n_M$. The MEC set $\hat{M}$ of size $n_M$ is equal to $M$, which is also the maximal MEC set. Since $\hat{\theta} =1$ and $\hat{n}_M = n_M$, both the estimated FLR and MMR are 0.

\paragraph{\textnormal{\emph{Case-M}}} Suppose $\gamma_{ab} = 0$ over $U$, and $\gamma_{ab} =1$ for $n_M -d$ pairs in $M$, while $\gamma_{ab} =0$ for $d$ pairs in $M$ due to key-variable errors. Again, $Q$ by (4) is not applicable. Using $Q_f$ by (3), we have $D_f = (n_M -d) \log \theta + d \log(1-\theta)$ and $\hat{\theta} = (n_M - d)/n_M$. By (9), we obtain $\hat{g}(1) = 1$ again, and $\hat{g}(0) = d/(n - n_M +d)$, so that the estimate of $n_M$ by (8) is $(n_M -d) +  d = n_M$. The maximal MEC set $M_1$ is of size $n_M -d$, whose estimated FLR and MMR are $0$ and $d/n_M$, respectively. Let $\hat{M}$ be a MEC set of size $n_M$. While $n_M - d$ pairs can be determined unequivocally as $M_1$, there are many possibilities for the remaining $d$ pairs among from $\Omega \setminus M_1$, all with $\gamma_{ab} = 0$. If $d_U$ of them belong to $U$, then both the FLR and MMR of $\hat{M}$ are $d_U/n_M$. Clearly, one may choose a MEC set with respect to the FLR and MMR, without being restricted to $|\hat{M}| = n_M$.

\paragraph{\textnormal{\emph{Case-U}}} Suppose $\gamma_{ab} = 1$ over $M$, and $\gamma_{ab} =1$ for $d$ pairs in $U$ due to key-variable errors, in such a manner that $n(1) = n_M + d$. Using $Q_f$ by (3), we obtain $\hat{\theta} =1$. We can estimate $n_M$ by solving (8) given $u(1) = d/(n- n_M)$, since $\hat{g}(0) = 0$ and 
\[
\hat{g}(1) = \frac{n_M/u(1)}{n_M \big( 1/u(1) -1 \big) + n} = \frac{n_M}{n_M - u(1) n_M + u(1) n} = \frac{n_M}{n_M + d}
\]
satisfy (8), i.e. $(n_M +d) n_M/(n_M +d) +  0 = n_M$. The maximal MEC set $M_1$ is of size $n_M +d$, whose estimated FLR and MMR are $d/(n_M + d)$ and $0$, respectively. For a MEC set of size $n_M$, one needs to exclude $d$ pairs from $M_1$. If $d_M$ of them belong to $M$, then both the FLR and MMR of the resulting $\hat{M}$ are $d_M/n_M$. Again, MEC may be guided by the error rates, without being restricted to $|\hat{M}| = n_M$.

\renewcommand{\theequation}{C.\arabic{equation}}
\setcounter{equation}{0}

\section*{C. Discussions}

\subsection*{C.1. Modelling comparison scores jointly}

Consider the possibility of MLE of $n_M$ and $m(\bgamma)$, where independent $\bgamma_{ab}$ are only assumed over the pairs in $M$, but not over $\Omega$ as in the classical approach. Let all the distinct \emph{entities} underlying the comparison space $\Omega$ be divided into three parts: the \emph{matched} entities $AB$, where each entity in $AB$ corresponds to a record pair in $M$; the \emph{unmatched} entities, each of which corresponds to, and is denoted by, either a record in $A_0 = \{ a\in A : \delta_a = 0\}$ or $B_0 = \{ b\in B : \delta_b = 0\}$, where $\delta_a = \sum_{b\in B} g_{ab}$ and $\delta_b = \sum_{a\in A} g_{ab}$. 

Assume $\bgamma_L = \{ \bgamma_M, \bz_{A_0}, \bz_{B_0} \}$ to be independent across the entities in $L = \{ AB, A_0, B_0\}$. Note that $(g_{\Omega}, \delta_A, \delta_B)$ are determined given $L$ and $\bgamma_L$ determined given $(L, \bz_A, \bz_B)$, where $\bz_A = \{ \bz_a : a\in A\}$ and $\bz_B = \{ \bz_b : b\in B\}$ contain the observed key vectors. Let $N_{AB} = |AB|$, $N_{A_0} = |A_0|$ and $N_{B_0} = |B_0|$, and $N_L = N_{AB} + N_{A_0} + N_{B_0}$. A model of the complete data $(\bz_A, \bz_B, L, N_L)$ can be given as
\[
f(\bz_A, \bz_B, L, N_L) = f(\bz_A, \bz_B | L) f(L | N_L) f(N_L)
\]
where $f(\bz_A, \bz_B | L)$ factorises over $AB$, $A_0$ and $B_0$, and $f(L | N_L)$ can be a multinomial model of $(N_{AB}, N_{A_0}, N_{B_0})$, and $f(N_L)$ a well-defined model of $N_L$ with its own parameters. The observed data are $(\bz_A, \bz_B)$ and $(n_A, n_B)$, where $n_A = N_{AB} + N_{A_0}$ and $n_B = N_{AB} + N_{B_0}$.

For instance, a model of  $(\bz_A, \bz_B)$ conditional on $L$ can be given as
\[
f(\bz_A, \bz_B | L) = \prod_{(a,b)\in \Omega} g_{ab} f_1(\bz_a, \bz_b; \btheta) ~ \prod_{a\in A} (1-\delta_a) f_0(\bz_a; \boldsymbol{\xi})  ~ \prod_{b\in B} (1 -\delta_b) f_0(\bz_b; \boldsymbol{\xi})
\]
where $f_1(\bz_a, \bz_b; \btheta)$ is the model of $(\bz_a, \bz_b)$ for a matched entity, and $f_0(\bz; \boldsymbol{\xi})$ is that of $\bz$ for an unmatched entity. Next, given $N_L = |L|$, a multinomial model of $L$ can be given as
\[
f(L | N_L) = \binom{N_L}{N_{AB}, N_{A_0}, N_{B_0}} \phi_{AB}^{N_{AB}} \phi_{A_0}^{N_{A_0}} \phi_{B_0}^{N_{B_0}} 
\]
where $\phi_{AB} + \phi_{A_0} + \phi_{B_0} =1$. Lastly, let $f(N_L)$ be a well-defined prior of $N_L$. 

Recall that the conditional expectation of $g_{ab}$ is calculated as $E(g_{ab} | \bgamma_{ab})$ in (8), marginally for a randomly selected record pair. However, for the E-step under the joint-data model here, one needs to evaluate $E(g_{ab} | \bz_A, \bz_B, n_A, n_B)$ conditional on all the observed data, which is intractable, since the conditional distribution of $(\bz_A, \bz_B, L, N_L)$ given $(\bz_A, \bz_B, n_A, n_B)$ does not factorise in any helpful way. It is also exceedingly impractical to calculate the observed likelihood based on $( \bz_A, \bz_B, n_A, n_B)$, which requires integrating $f(\bz_A, \bz_B, L, N_L)$ over all $(L, N_{AB}, N_{A_0}, N_{B_0})$ that are compatible with the observed $(n_A, n_B)$.

Therefore, MLE by modelling the comparison scores jointly does not seem viable.

\subsection*{C.2 Moment-based estimation}

It is possible to derive a moment-based estimator of the perturbation probability $\alpha_k$, for $k=1, ..., K$. Let $\xi_k = \mbox{Pr}(\gamma_{ab,k} =1 | g_{ab} =0)$ for a randomly selected record pair from $\Omega$. Under the non-informative hit-miss model, where $\beta_{kd} = m_{kd} = u_{kd}$, we have
\[
\xi_k = \sum_{d=1}^{D_k} \beta_{kd}^2 \qquad\text{and}\qquad \theta_k =1 - \alpha_k (1 - \xi_k)
\]
Let $n(1;k)$ be the observed number of pairs that agree on $z_k$, and let $n(0;k) = n - n(1;k)$.
It follows from $E\big( n(1; k) | n_M \big) = n_M \theta_k + (n - n_M) \xi_k$ that  
\begin{equation} \label{alpha}
n_M (1 - \alpha_k) (1 - \xi_k) = n (1 - \xi_k) - E\big( n(0; k) | n_M \big) 
\end{equation}
Plugging in the estimates of $\xi_k$ and $n_M$, and $n(0;k)$ for its expectation, we obtain an estimate of $\alpha_k$, subjected to the assumption that $\alpha_k$ is positive for some $k=1, ..., K$. 

Moreover, setting $\alpha_k =0$ in \eqref{alpha} yields an approximate estimating equation for $n_M$,  
\[
\Delta_k = n(0; k) - (n - n_M) \big( 1- \hat{\xi}_k \big) = 0 
\quad\text{subjected to}\quad n_M \leq \min(n_A, n_B)
\]
Note that in principle this makes it possible to estimate $n_M$ based on a single binary key variable, which represents the minimal setting at all conceivable for record linkage. For $K>1$, a possibility is to use the \emph{minimum distance estimator} \citep{wolfowitz1957minimum} of $n_M$, which is given by the value that minimises a chosen metric that weighs together all the $\Delta_k$, where $k=1, ..., K$. Note that the bias of the resulting $\hat{n}_M$ does not vanish as $n\rightarrow \infty$, as long as some $\theta_k$ converges to a value unequal to 1. 

However, the main problem with these moment-based estimates is that they are very unstable. This is largely caused by the fact that the bipartition of the comparison space $\Omega = M \cup U$ is highly imbalanced, where $\pi = n_M/n$ is very close to 0, such that the right-hand side of \eqref{alpha} is the difference between two numbers, both of which are much larger than the term on the left-hand side of the equation.

The moment equation \eqref{alpha} can also be related to the objective function (2). Let $n_M(1;k)$ and $n_U(1;k)$ be, respectively, the number of $\gamma_k =1$ in $M$ and $U$, and $n_M(0;k)$ and $n_U(0;k)$ those of $\gamma_k =0$. Assume the model (5) of $m(\bgamma; \btheta)$, and similarly of $u(\bgamma; \boldsymbol{\xi})$. The score functions derived from the objective function (2) are given by $\pi = n_M/n$, and  
\begin{gather*}
\begin{cases} \frac{\partial h}{\partial \theta_k} = \frac{n_M(1;k)}{\theta_k} - \frac{n_M(0;k)}{1-\theta_k} = 0 \\
\frac{\partial h}{\partial \xi_k} = \frac{n_U(1;k)}{\xi_k} - \frac{n_U(0;k)}{1-\xi_k} = 0 \end{cases}
\Leftrightarrow\quad
\begin{cases} n_M \theta_k = n_M(1;k) = n(1;k) - n_U(1;k) \\ (n- n_M) \xi_k = n_U(1;k) = n(1;k) - n_M(1;k) \end{cases} \\
\Leftrightarrow\quad 
\begin{cases} n(0; k) = n_M (1- \theta_k) + n_U(0; k) \\ n(0; k) = n_M(0;k) + (n- n_M) (1-\xi_k) \end{cases}
\end{gather*}
for $k=1, ..., K$. Taking expectation of both score functions, we recover \eqref{alpha} above.  
In other words, had one observed the missing data $g_{\Omega}$ and maximised $h(\boldsymbol{\eta})$ by (2) directly, as if it were the complete-data log-likelihood based on $(g_{\Omega}, \gamma_{\Omega})$, one would have been conducting rather inefficient moment-based estimation instead. 

Finally, note that the imbalanced bipartition of $\Omega$, which affects \eqref{alpha}, has no bearing on the unsupervised MEC algorithm, where one uses a value of $\theta_k$ that satisfies (10), which depends on two numbers that are of the same magnitude of $n_M$.

\renewcommand{\theequation}{D.\arabic{equation}}
\setcounter{equation}{0}

\section*{D. Additional simulations with low levels of overlap}

Another simulation study was performed to assess the performance of the proposed method when the degree of files' overlap is small. Table 6 and Table 7 present the results obtained for $p_A = 0.2, 0.15,$ and $0.1$ (i.e., $\pi=0.0002, 0.00015,$ and $0.0001$) under Scenario I and Scenario II, respectively. As expected, the the estimate $\hat{n}_M$ (or $\hat{\pi}$) gets worse as $p_A$ (or $\pi$) reduces, which is consistent with the previous findings of \citet{sadinle2017bayesian} and \citet{enamorado2019using}. 

\begin{table}[ht]
	\centering
	\caption{Parameters and averages of their estimates, averages of error rates and their estimates, over 200 simulations under Scenario I. Median of estimate of $n_M$ given as $\tilde{n}_M$.} 
	\label{tb1}
	\begin{tabular}{cc|cc|ccccccc} \hline
		\multicolumn{2}{c|}{Parameter} & \multicolumn{2}{c|}{Formulae} & \multicolumn{7}{c}{Estimation} \\ \cline{1-2}\cline{3-4}\cline{5-11}
		$\pi$ & $E(n_M)$ & $\theta_k^{(t)}$ & $\xi_k^{(t)}$ & $\hat{\pi}$ & $\hat{n}_M$ & $\tilde{n}_M$ & $\mbox{FLR}$ & $\mbox{MMR}$ & $\widehat{\mbox{FLR}}$ & $\widehat{\mbox{MMR}}$\\ \hline
		\multirow{4}{*}{{.0002}} & \multirow{4}{*}{{100}} & $\hat{\theta}_k$ &  (15)  
		& .00020 & 100.7 & 100 & .0460 & .0395 & .0382 & .0382 \\
		& & (13) & (20) & .00026 & 132.0 & 107 & .1434 & .0400 & .0983 & .0983 \\
		& & (13) & (15) & .00033 & 162.9 & 110 & .2040 & .0358 & .1417 & .1417 \\
		&  & (18) & (15) &.00077 & 387.4  & 500 & .6204 & .1406 & .3023 & .3023 \\  \hline
		\multirow{4}{*}{{.00015}} & \multirow{4}{*}{{75}} & $\hat{\theta}_k$ &  (15)  
		& .00015 & 75.3 & 75 & .0495 & .430 & .0384 & .0384 \\
		& & (13) & (20) & .00024 & 117.9 & 81 & .1699 & .0457 & .1125 & .1125 \\
		& & (13) & (15) & .00031 & 152.9 & 84 & .2420 & .0412 & .1631 & .1631 \\
		&  & (18) & (15) & .00084 & 420.3 & 500 & .7250 & .1579  & .3496  & .3496 \\  \hline
		\multirow{4}{*}{{.0001}} & \multirow{4}{*}{{50}} & $\hat{\theta}_k$ &  (15)  
		& .00010 & 50.2 & 50 & .0582 & .0520 & .0397 & .0397 \\
		& & (13) & (20) & .00022 & 109.6 & 55 & .2163 & .0529 & .1365 & .1365 \\
		& & (13) & (15) & .00031 & 155.1 & 57 & .3030 & .0485 & .1979 & .1979 \\
		&  & (18) & (15) & .00092 & 459.8 & 500 & .8424  & .1736  & .4015 & .4015\\  \hline
	\end{tabular}
\end{table}

\begin{table}[ht]
	\centering
	\caption{Parameters and averages of their estimates, averages of error rates and their estimates, over 200 simulations under Scenario II. Median of estimate of $n_M$ given as $\tilde{n}_M$.} 
	\label{tb2}
	\begin{tabular}{cc|cc|ccccccc} \hline
		\multicolumn{2}{c|}{Parameter} & \multicolumn{2}{c|}{Formulae} & \multicolumn{7}{c}{Estimation} \\ \cline{1-2}\cline{3-4}\cline{5-11}
		$\pi$ & $n_M$ & $\theta_k^{(t)}$ & $\xi_k^{(t)}$ & $\hat{\pi}$ & $\hat{n}_M$ & $\tilde{n}_M$ & $\mbox{FLR}$ & $\mbox{MMR}$ & $\widehat{\mbox{FLR}}$ & $\widehat{\mbox{MMR}}$\\ \hline
		\multirow{4}{*}{{.0002}} & \multirow{4}{*}{{100}} & $\hat{\theta}_k$ & (15)
		& .00020 & 100.7 & 100 & .0457 & .0394 & .0369 & .0369 \\
		& & (13) & (20) & .00022 & 111.0 & 103 & .0856 & .0436 & .0592 & .0592 \\
		& & (13) &  (15) & .00024 & 118.7 & 104 & .1102 & .0418 & .0740 & .0740 \\
		&  & (18) & (15) & .00050 & 249.0 & 110 & .3583  & .0923  & .1875  & .1875  \\  \hline
		\multirow{4}{*}{{.00015}} & \multirow{4}{*}{{75}} & $\hat{\theta}_k$ & (15) 
		& .00015 & 75.7 & 75 & .0496 & .0418 & .0367 & .0367 \\
		& & (13) & (20) & .00017 & 86.3 & 78 & .0952 & .0461 & .0600 & .0600 \\
		& & (13) &  (15) & .00021 & 106.4 & 79 & .1424 & .0442 & .0920 & .0920 \\
		&  & (18) & (15) & .00060  & 301.6  & 500 & .5031  & .1193  & .2528  & .2528  \\  \hline
		\multirow{4}{*}{{.0001}} & \multirow{4}{*}{{50}} & $\hat{\theta}_k$ & (15) 
		& .00010 & 50.5 & 50 & .0528 & .0445 & .0377 & .0377 \\
		& & (13) & (20) & .00013 & 63.6 & 52 & .1111 & .0501 & .0649 & .0649 \\
		& & (13) &  (15) & .00016 & 82.3 & 53 & .1567& .0478 & .0945 & .0945 \\
		&  & (18) & (15) & .00077 & 385.5 & 500 &.7034 & .1537 & .3564 & .3564 \\  \hline 
	\end{tabular}
\end{table}

\bibliographystyle{chicago}
\bibliography{RefSMJ}

\end{document}